\documentclass[prd,nofootinbib,tightenlines]{revtex4}
\usepackage[dvips]{graphicx}
\def\journal#1, #2, #3#4, #5#6#7#8    {
    {#1~} {#2}  (#5#6#7#8) #3#4}

\usepackage{amssymb}
\usepackage{amsmath,amsfonts,amssymb,amscd,enumerate}
\usepackage{amsmath}
\usepackage{lscape}
\draft
\begin{document}


\renewcommand{\thesection}{\arabic{section}}
\renewcommand{\thesubsection}{\thesection.\arabic{subsection}}
\renewcommand{\theequation}{\arabic{equation}}
\renewcommand {\c}  {\'{c}}
\newcommand {\cc} {\v{c}}
\newcommand {\s}  {\v{s}}
\newcommand {\CC} {\v{C}}
\newcommand {\C}  {\'{C}}
\newcommand {\Z}  {\v{Z}}
\newcommand{\pv}[1]{{-  \hspace {-4.0mm} #1}}

\newcommand{\be}{\begin{equation}} \newcommand{\ee}{\end{equation}}
\newcommand{\bea}{\begin{eqnarray}}\newcommand{\eea}{\end{eqnarray}}
\newcommand{\grad}{\bm \nabla}

\def\A{\mathcal{A}}
\def\C{\mathbb{C}}
\def\N{\mathbb{N}}
\def\R{\mathbb{R}}
\def\Z{\mathbb{Z}}
\def\g{\mathfrak{g}}
\def\h{\mathfrak{h}}
\def\p{\partial}
\def\d{\delta}
\def\proof{\noindent \textbf{Proof. }}
\def\qed{$\blacksquare$}
\def\F{\mathcal{F}}
\def\G{\mathcal{G}}
\def\H{\mathcal{H}}
\def\K{\mathcal{K}}
\def\x{\hat x}

\baselineskip=14pt

\begin{center}
{\bf  \Large  Scalar field theory on $\kappa$-Minkowski spacetime and \\
 translation and  Lorentz invariance} 
\bigskip
 
S. Meljanac {\footnote{e-mail: meljanac@irb.hr}} and
 A. Samsarov {\footnote{e-mail: asamsarov@irb.hr}} \\  
 Rudjer Bo\v{s}kovi\'c Institute, Bijeni\v cka  c.54, HR-10002 Zagreb,
Croatia \\[3mm]

\end{center}
\setcounter{page}{1}
\bigskip


\begin{abstract}
   We investigate the
properties of $\kappa$-Minkowski spacetime by using  representations
of the corresponding deformed algebra
    in terms of undeformed Heisenberg-Weyl algebra. The deformed
    algebra consists of $\kappa$-Poincar\'{e} algebra extended with
    the generators of the deformed Weyl algebra.
The part of deformed algebra, generated by rotation, boost and
momentum generators, is described by the Hopf algebra structure.
   The approach used in our considerations is completely Lorentz covariant.
   We further use an adventages of this approach to consistently construct
   a star product which has a property that under
   integration sign it can be replaced
   by a standard  pointwise multiplication, a property that was since
   known to hold for Moyal, but not also for $\kappa$-Minkowski spacetime.
 This star product also has generalized
trace and cyclic properties and the construction alone is accomplished
  by considering a classical Dirac operator representation of 
  deformed algebra and by requiring it to be hermitian. We find that
  the obtained star product is not translationally invariant, leading to a
  conclusion that the classical Dirac operator representation is the
  one where translation invariance cannot simultaneously be
  implemented along with hermiticity. 
  However, due to the integral property satisfied by the star product,
 noncommutative free scalar field theory does not have a problem with
 translation symmetry breaking and can be shown to
  reduce to an ordinary free scalar field theory without nonlocal
  features and tachionic modes and basicaly
   of the very same form. The issue of Lorentz invariance of the theory is also discussed.

\bigskip 
\noindent
PACS number(s): 03.65.Fd, 11.10.Nx, 11.30.Cp 
\end{abstract}


\maketitle
 


\section{Introduction}

There are two basic areas of interest which make $\kappa$-Minkowski spacetime 
especially interesting object of theoretical investigation from both,
physical as well as mathematical perspective. Since it emerges naturally from 
$\kappa$-Poincar\'{e} algebra \cite{Lukierski:1991pn},\cite{Lukierski:1993wx},
 which provides a group theoretical framework for
describing symmetry lying in the core of the Doubly Special Relativity
(DSR) theories \cite{AmelinoCamelia:2000ge},\cite{AmelinoCamelia:2000mn},\cite{Magueijo:2001cr},\cite{Magueijo},
 it is thus convenient spacetime candidate for DSR theories.
Although different proposals for DSR theories can be looked upon as different
bases \cite{KowalskiGlikman:2002we},\cite{KowalskiGlikman:2002jr}
 for $\kappa$-Poincar\'{e} algebra, they all have in their core
the very same noncommutative structure, encoded within
$\kappa$-deformed algebra. Other important argument that favours $\kappa$-Minkowski spacetime 
is the strong indication that it as well arises in the context of
quantum gravity coupled to matter fields \cite{AmelinoCamelia:2003xp},\cite{Freidel:2003sp}. These considerations
show that after integrating out topological degrees of freedom of
gravity, the effective dynamics of matter fields is described by a
noncommutative quantum field theory which has a $\kappa$-Poincar\'{e}
group as its symmetry group \cite{Freidel:2005bb},\cite{Freidel:2005me},\cite{Freidel:2005ec}. 
In this context
the $\kappa$-Minkowski can be, and there are some arguments to
support this,
 considered as a flat limit of quantum gravity in pretty much
the same way as the special relativity corresponds to general
relativity when the same limit is concerned.

This situation makes a noncommutative field theories on noncommutative spacetimes,
particularly those of the $\kappa$-Minkowski type, even more interesting
subject to study. Various attempts have been undertaken towards this
direction by many authors
 \cite{Kosinski:1999ix},\cite{Kosinski:1999dw},\cite{AmelinoCamelia:2001fd},\cite{Daszkiewicz:2004xy},\cite{Dimitrijevic:2003wv},
including various possibilities for constructing field theories
 on $\kappa$-Minkowski spacetime and investigating their properties.
Recently it is well established that if 
$\kappa$-Poincar\'{e} Hopf algebra is supposed to be a plausible model
for describing physics in $\kappa$-Minkowski spacetime, then it is
necessary to accept certain modifications in statistics obeyed by the
particles. This means that $\kappa$-Minkowski spacetime leads to
modification of particle statistics which results in deformed
oscillator algebras
 \cite{gov},\cite{Arzano:2007ef},\cite{Daszkiewicz:2007az},\cite{luk1},\cite{luk2},\cite{Young:2007ag},\cite{Arzano:2008bt}.
Deformation quantization of Poincar\'{e} algebra can be performed by means of the twist operator
\cite{drinfeld},\cite{Borowiec:2004xj},\cite{borluk} which happens to
include dilatation generator, thus belonging to the universal
enveloping algebra of the general linear algebra
 \cite{Govindarajan:2008qa},\cite{Bu:2006dm},\cite{Borowiec:2008uj},\cite{Kim:2008mp},\cite{Kim:glavni},\cite{Borowiec:2010yw}.
This twist operator gives rise to a deformed statistics on
$\kappa$-Minkowski spacetime \cite{Govindarajan:2008qa},\cite{Young:2007ag},\cite{Young:2008zg},\cite{Young:2008zm}.
In correspondence to these observations, bounds can be put on the
quantum gravity scale by using deformed statistics results, in the
context of atomic physics \cite{Harikumar:2009wv},\cite{Arzano:2009bd} as well as by comparing deformed dispersion
relations to corresponding time delay calculations of high energy photons
\cite{Borowiec:2009ty}.

During last decade or so there has been quite large effort made to
formulate integration on $\kappa$-Minkowski and to find a relevant
meassure that would consequently be  used in formulating field theory.
The issue of proper definition of the
integral on $\kappa$-Minkowski spacetime, that would have desired
trace and cyclic properties, was considered by various groups of
authors.  In
Refs.\cite{Felder:2000nc},\cite{Dimitrijevic:2003wv},\cite{Dimitrijevic:2003pn},\cite{Moller:2004sk},\cite{Agostini:2006zza}
the problem of cyclicity of the integral is tackled in a way that
required the introduction of a new integration meassure which then had
to satisfy certain conditions, stated in a form of differential
equations. On the other hand, in Ref.\cite{Chryssomalakos:2007jr} the authors
 were considering a star
product for noncommutative spaces of Lie type, which
provides a definition of an invariant integral, satisfying
quasicyclicity property. According to their claims, this
quasicyclicity property reduces to exact cyclicity in the case when
the adjoint representation of the underlying Lie algebra is traceless.
 Several other attempts
for identifying a proper integral identity that would enclose
most of peculiarities of $\kappa$-Minkowski have been made. In Ref.\cite{Freidel:2006gc}
an example of such integral identity is obtained which has an
adventage that it does not require a modification of an integration
meassure, but deals with a star product that leads to unphysical modes,
such as the appeareance of tachyons. The appeareance of tachyonic modes
draws its origin in basicaly nonlocal character of the action
obtained in \cite{Freidel:2006gc}.
 The same identity is again mentioned in \cite{KowalskiGlikman:2009zu}, and a similar one is
redireved in Ref.\cite{Meljanac:2007xb} in a more general setting corresponding to a
general class of various orderings.

In the present paper we are further investigating the basic mathematical properties of 
$\kappa$-Minkowski spacetime and consequences they have on the
physics, especially on the matters regarding nonlocality and
translation and Lorentz invariance.
Particularly we shall show that it is
possible to introduce a star product in $\kappa$-Minkowski spacetime,
that,  under the integration sign, can be replaced by the standard pointwise multiplication.
The procedure for showing this is governed by simple physical principles and
appears to be internally consistent.
The same star product also happens to have a generalized trace and cyclic
properties that can reduce to standard trace and cyclic properties,
 if certain conditions, imposed on physical fields, are satisfied.
One additional point that is specific about our approach is that the
issue with meassure on $\kappa$-Minkowski is in a sort of way avoided by absorbing it within
a new star product in such a way that the new star product has a correct limit when
the parameter of deformation goes to zero.
 All this can be shown within our approach that
uses a method of realizations. In this method the generators
of deformed algebra, which include noncommutative coordinates and 
generators of the Poincar\'{e} algebra, are represented in terms of generators of
the undeformed Heisenberg algebra, i.e. in terms of ordinary
coordinates and their respective derivatives. The important point
about this procedure is
that  from the beginning, we start with the integral on
$\kappa$-Minkowski having ordinary integration meassure
and demand that realization of deformed algebra be hermitian. The stated
requirement shows up as a crucial step, which allows for the
construction of the star product, whose generalized trace and cyclic properties
then emerge as a beneficial side effects. In this way
 the generalized trace property of the integral on
$\kappa$-Minkowski spacetime arises naturally, simply by requiring from the realization of
deformed algebra to be hermitian. Though, the star product resulting
from the hermitization procedure appears to be translationally
noninvariant. This feature shows that in the classical Dirac operator
representation, the one in which we are making our analysis, hermiticity and translation invariance mutually
interfere in a sense that it is not possible to simultaneously have
both of these properties satisfied on $\kappa$-Minkowski, at least as
far as the classical Dirac operator representation is concerned.

The symmetry underlying $\kappa$-Minkowski spacetime is described
by the $\kappa$-deformed Poincar\'{e} algebra, whose mathematical
structure is most conveniently specified with Hopf algebra. The investigations here
will be pursued in a so called classical basis \cite{KowalskiGlikman:2002we},\cite{KowalskiGlikman:2002jr},\cite{Meljanac:2007xb},\cite{Borowiec:2009vb},\cite{KresicJuric:2007nh} where the algebraic
part of $\kappa$-Poincar\'{e} Hopf algebra is undeformed and all
deformations are contained within the coalgebraic sector. This  means
that the action of Poincar\'{e} generators on $\kappa$-Minkowski
spacetime and consequently on the algebra of
noncommutative functions will change with respect to their action on
ordinary Minkowski space, giving rise to the appropriate
 modifications of Leibniz rules and corresponding coproducts.
The approach followed in this paper is developed and widely elaborated
in a series of papers
\cite{Meljanac:2007xb},\cite{KresicJuric:2007nh},\cite{Meljanac:2006ui},
\cite{Durov:2006iv},\cite{Meljanac:2008ud},\cite{Meljanac:2008pn},\cite{Meljanac:2010qp}
 that are concerned with classifying of noncommutative spaces and
investigations of various properties of their realizations. The general
principle established there is that to each realization of
noncommutative space, there corresponds a particular ordering prescription
and definite coproduct and star product as well as a twist operator.
Here we shall use one particular realization, the so called classical
Dirac operator representation, that in this series of
papers is referred to as the  
  {\it{natural realization}} and which is characterized by the simple
  requirement that the deformed and undeformed derivatives have to be the
  same, i.e. identified.
   The classical Dirac operator representation was explicitly or
   implicitly considered by many authors
 \cite{Maggiore:1993rv},\cite{Maggiore},\cite{Dimitrijevic:2003wv},\cite{Dimitrijevic:2003pn},\cite{Moller:2004sk},\cite{Freidel:2006gc}
 and it seems to be of special
   importance to physics since it can be related to a generalized
   uncertainty principle that has its origin in the study of high
   energy collisions of strings \cite{Gross:1987ar}. It is thus also directly related to
   the existence of a minimal length \cite{Kempf:1994su},\cite{Kempf:1996nk}.
 For this classical Dirac operator
  representation we find a corresponding star product. 
After requiring this realization to be hermitian, we arrive
at the result that, under the
integration sign, the corresponding star product
can be replaced with the pointwise multiplication  and also
 happens to satisfy a generalized trace property.
   The most of methods used in investigations that were performed here are taken over from the
  Fock space analysis carried out in \cite{multimode},\cite{perinvalg},\cite{bardek},\cite{collapse},\cite{bosonic},\cite{Jonke:2002kb},\cite{acta}.

The plan of paper is the following. In section 2 we introduce an
algebraic setting we are working in. It includes the algebra of
coordinates on $\kappa$-Minkowski spacetime, an algebra of symmetry
generators, which we take to be a undeformed Poincar\'{e} algebra,
and the action of symmetry generators on $\kappa$-Minkowski spacetime.
It also includes notion of a shift operator and a definition of
the star product that we shall use throughout this paper.
In this section we also introduce a classical Dirac operator
representation of the $\kappa$-deformed algebra and the star product that
we  use is defined with respect to this representation.
 In section 3 we further elaborate on the algebraic
properties of $\kappa$-Minkowski spacetime and find that these properties are
enclosed within a Hopf algebra structure. In particular we list the
coproducts and antipodes of $\kappa$-Poincar\'{e} symmetry generators.
Section 4 is devoted to analyzing the properties of the star product
corresponding to classical Dirac operator representation ({\it{natural realization}})
and an integral identity satisfied by the star product is there rederived in
a new way, exposing its nonlocal properties.
This star product, before hermitization is carried out, is shown to be
translationally invariant.
 The process of
hermitization of the classical Dirac operator representation
is then carried out in section 5 and is found to lead to a
cancellation of nonlocal operators appearing in the theory, giving
rise to generalized trace and cyclic properties of the star
product. The hermitization procedure also
enables a proper and consistent introduction of hermitian conjugation 
(i.e. adjoint operation) and makes a correspondence between commutative
and noncommutative algebras of functions fully established. However,
the star product that is obtained in this way, as a result of
hermitization, is not translationally invariant. In the conclusion we further discuss
this isssue of translational noninvariance of the hermitized star
product and
 finish the paper with discussion regarding the problems with Lorentz invariance of the theory and
conclude by suggesting a proposal about a way in which Lorentz symmetry can be restored.

\section{$\kappa$-Minkowski spacetime}

We consider a $\kappa$-deformed
  Minkowski spacetime whose noncommutative
coordinates $\x_{\mu}, ~ (\mu = 0,1,...,n-1),$ close a Lie algebra
with the Lorentz generators $M_{\mu\nu},~ (M_{\mu \nu} = -M_{\nu \mu})
$,
\begin{align} \label{2.1}
 [\x_{\mu},\x_{\nu}] & = i(a_{\mu}\x_{\nu}-a_{\nu}\x_{\mu}), \\ 
 [M_{\mu \nu}, M_{\lambda \rho}]  & =  \eta_{\nu \lambda}M_{\mu \rho} -
 \eta_{\mu \lambda}M_{\nu \rho}
 -\eta_{\nu \rho} M_{\mu \lambda} + \eta_{\mu \rho} M_{\nu \lambda}, \label{2.2} \\
 [M_{\mu\nu}, \x_{\lambda}] & = \x_{\mu} \eta_{\nu\lambda} - \x_{\nu}
  \eta_{\mu\lambda}-i\left( a_{\mu}  M_{\nu\lambda}-a_{\nu} 
  M_{\mu\lambda} \right), \label{2.3}
\end{align}
where $a_{\mu}$ are components of the deformation
  vector and $~ \eta_{\mu\nu} = diag(-1,1,\cdot \cdot \cdot, 1)$
defines a spacetime metric. The quantity $a^2 = a_{\mu}a^{\mu}$ 
is Lorentz invariant having a dimension of inverse mass squared,
 $a^2 \equiv \frac{1}{{\kappa}^{2}}.$ In the smooth limit $a_{\mu} \rightarrow 0,$ the above algebra reduces to a
 standard one, describing commutative spacetime.
 It is easy to check that all Jacobi identities for this algebra are satisfied.
 Throughout the paper we
shall work in units  $\hbar = c =1.$ 

The symmetry of the deformed spacetime (\ref{2.1}) is assumed to be
described by an undeformed Poincar\'{e}
algebra. Thus, in addition to Lorentz generators  $M_{\mu\nu}, $ we
also introduce momenta $P_{\mu}$ which transform as vectors under the
Lorentz algebra, 
\begin{align} \label{2.4}
[P_{\mu},P_{\nu}]&=0,  \\
[M_{\mu\nu},P_{\lambda}]&= \eta_{\nu\lambda}\,
P_{\mu}-\eta_{\mu\lambda}\, P_{\nu}. \label{2.5}
\end{align}
 The algebra (\ref{2.1})-(\ref{2.5}), however, does not fix the commutation
relation between $~P_{\mu}$ and $\x_{\nu}$. In fact, there are
infinitely many possibilities for the commutation
relation between $~P_{\mu}$ and $\x_{\nu},$ all of which are consistent
with the algebra (\ref{2.1})-(\ref{2.5}) in a sense that Jacobi
identities are satisfied between all generators of the algebra.   
In this way we have  an
extended algebra, which includes generators $M_{\mu\nu}, ~P_{\mu}$ and $\x_{\lambda},$
 and has Jacobi identities satisfied for all
combinations of the generators $M_{\mu\nu},$ $P_{\mu}$ and $\x_{\lambda}.$
Particularly, the algebra generated by $P_{\mu}$ and $\x_{\nu}$ is a deformed
Heisenberg-Weyl algebra which can generally be written in the form
\begin{equation} \label{2.6}
[P_{\mu},\x_{\nu}] = -i \Phi_{\mu\nu}(P),
\end{equation}
where $ \Phi_{\mu\nu}(P)$ are functions of generators $P_{\mu},$
 satisfying the boundary conditions
$ \Phi_{\mu\nu}(0)=\eta_{\mu\nu}$ and consistent with relation
 (\ref{2.1}) and Jacobi identities that have to be fulfilled for all
 combinations of generators $M_{\mu\nu},$ $P_{\mu}$ and $\x_{\lambda}.$

The momentum $P_{\mu} = -i \hat{\p}_{\mu},$ expressed in terms
 of deformed derivative $ \hat{\p}_{\mu},$ can be realized in a
 natural way \cite{Meljanac:2007xb} by adopting the identification between deformed and
 undeformed derivatives, $ \hat{\p}_{\mu} \equiv \p_{\mu},$ implying
 $P_{\mu} = -i \p_{\mu}. $
The deformed algebra (\ref{2.1})-(\ref{2.5}) then admits a wide class of realizations
\begin{equation} \label{2.7}
\x_{\mu} =x^\alpha \Phi_{\alpha\mu}(P),
\end{equation}
\begin{equation} \label{2.8}
M_{\mu\nu} = x_{\mu}\p_{\nu}- x_{\nu}\p_{\mu}, 
\end{equation}
in terms of the generators of the undeformed Heisenberg algebra,
\begin{equation} \label{2.9}
[x_{\mu},x_{\nu}]=0, \qquad  [\p_{\mu},\p_{\nu}]=0, \qquad  [\p_{\mu},x_{\nu}] = \eta_{\mu\nu}
\end{equation}
and analytic function $\Phi_{\alpha\mu}(P),$
appearing on the right hand side of Eq.(\ref{2.6}).
By taking this prescription, the deformed algebra
(\ref{2.1})-(\ref{2.5}) is then automatically satisfied, as
well as all Jacobi identities among $\x_\mu$, $M_{\mu\nu},$ and
$P_{\mu}.$ 
We take one particular realization from the class (\ref{2.7}), namely
the one of the form
\begin{equation} \label{2.10}
\x_{\mu}=x_{\mu}(-ia_{\alpha}{\p}^{\alpha}+\sqrt{1-a^2{\p}_{\alpha}{\p}^{\alpha}})+i(ax){\p}_{\mu}.
\end{equation}
This particular realization of NC spacetime (\ref{2.1}) is usually
 known as the
 classical Dirac operator representation and has been
considered for the first time in Ref.\cite{Dimitrijevic:2003wv}.
It is consistent with the deformed algebra
(\ref{2.1})-(\ref{2.5}) and is also a special case of 
the whole family of realizations of NC spacetime algebra (\ref{2.1}) considered in \cite{Meljanac:2009fy},
 where it is classified as the Maggiore-type of realizations \cite{Maggiore:1993rv},\cite{Maggiore}. In
 Ref.\cite{Meljanac:2007xb} it is also refered to as {\it{natural
 realization}}. Throughout the paper we shall  work only with this
 type of realization and later on (section 5) with its hermitian form.

With these particular settings, the deformed Heisenberg-Weyl
algebra (\ref{2.6}) looks as 
\begin{align} \label{2.12}
[P_{\mu},\x_{\nu}]=-i\eta_{\mu\nu} \left(aP+\sqrt{1+ a^2 P^2}
  \right) +i a_\mu
P_\nu.
\end{align}
There also exists a
universal shift operator $ Z^{-1}$ \cite{KresicJuric:2007nh} with the following properties:
\begin{equation} \label{2.13}
[Z^{-1},\x_{\mu}] = -ia_{\mu} Z^{-1}, \qquad [Z,P_{\mu}] =0,
\end{equation}
where $Z$ is a regular operator, $Z Z^{-1} = 1. $
As an implication of  these two equations we have
\begin{equation} \label{2.15}
 [Z,\x_{\mu}] = ia_{\mu} Z, \qquad \x_{\mu}Z\x_{\nu} =\x_{\nu}Z\x_{\mu}.
\end{equation}
The explicit realization of the universal shift operator $Z^{-1}$ 
in terms of generators ${\p}_{\mu}$ of the Weyl algebra has the form
\begin{equation} \label{16}
  Z^{-1} = -ia_{\alpha}{\p}^{\alpha} + \sqrt{1 - a^2{\p}_{\alpha}{\p}^{\alpha}}.
\end{equation}
As a consequence, the Lorentz generators can be expressed in terms of
$Z$ as
\begin{equation} \label{2.17}
M_{\mu\nu}=i(\x_{\mu}P_{\nu}-\x_{\nu}P_{\mu})Z,
\end{equation}
and one can also show the validity of the relation
\begin{equation} \label{2.18}
[Z^{-1},M_{\mu\nu}] = a_{\mu}P_{\nu}-a_{\nu}P_{\mu}.
\end{equation}

It is to expect that a deformation of the spacetime
structure will affect the algebra of physical fields, leading to a
modification of multiplication in the corresponding universal
enveloping algebra. Specifically, it means that a spacetime deformation
requires a replacement of the usual pointwise multiplication with a
deformed product or star product. This will consequently have
   an impact on physics,
particularly it will modify the way in which the field theoretic
action should be constructed. It is for this reason that
 we introduce a  star product in a given realization
 (\ref{2.10}). This star product can be introduced in the
following way. First
 we define the unit element $1$ as
\begin{equation} \label{2.19}
\begin{array}{c}
  \partial_{\mu} \triangleright 1 ~ = ~0, \qquad  M_{\mu\nu}
  \triangleright 1 ~ = ~ 0.
\end{array}
\end{equation}
This means that Poincar\'{e} generators give zero, when acting on unit
element $1$.

Then, for the particular realization of noncommutative spacetime
(\ref{2.1}), given by (\ref{2.10}), there is a unique map from the
algebra ${\mathcal{A}}_{a}$ of fields $\phi(x)$ in commutative
coordinates $x_{\mu}$ to the enveloping algebra ${\hat{\mathcal{A}}}_{a}$ of
noncommutative fields $ \hat{\phi}(\hat{x})$ in NC coordinates
$\hat{x}_{\mu}$. This map $\; \Omega : ~ {\mathcal{A}}_{a} \longrightarrow
{\hat{\mathcal{A}}}_{a} \;$ can uniquely be characterized by
\begin{equation} \label{2.20}
 \Omega : ~ {\mathcal{A}}_{a} \longrightarrow {\hat{\mathcal{A}}}_{a}, \quad
  \phi(x) \longmapsto \hat{\phi}(\hat{x}) \quad
\mbox{such that}  \quad  \hat{\phi}(\hat{x})  \triangleright 1 ~ = ~ \phi(x).
\end{equation}
If we further have two fields  $ \hat{\phi}(\hat{x}),  \hat{\psi}(\hat{x})$ in NC coordinates, make their
product $ \hat{\phi}(\hat{x})  \hat{\psi}(\hat{x})$ and ask
 which field in commutative coordinates this
combination belongs to (through the mapping $\; \Omega \;$), we arrive at
\begin{equation} \label{2.21}
(\phi \star \psi)(x) =
 \hat{\phi}(\hat{x}) \hat{\psi}(\hat{x}) \triangleright 1 ~ = ~
 \hat{\phi}(\hat{x}) \triangleright \psi(x).
\end{equation}
This prescription defines the star product in any realization and thus
 specifically defines the star product in the realization (\ref{2.10}), too.
In Eq.(\ref{2.21}) it is assumed that $\hat{\phi}(\hat{x})  \triangleright 1 ~ = ~ \phi(x)$
and $\hat{\psi}(\hat{x})  \triangleright 1 ~ = ~ \psi(x)$ and that
 $\x$ is given by (\ref{2.10}). It is also assumed that
 $\hat{\phi}(\hat{x})$ and $\hat{\psi}(\hat{x})$ are functions of NC
 coordinates only and not of derivatives.

    In the setting just described, we are considering the fields $ \hat{\phi}(\hat{x})$
formed out of polynomials in $\hat{x}$, which constitute an universal
enveloping algebra ${\hat{\mathcal{A}}}_{a}$ in $\hat{x},$ with the
observation that
  in the algebra ${\hat{\mathcal{A}}}_{a}$ the multiplication is given
  by the standard composition of the operators.
  In the similar way fields $\phi(x)$ in
commutative coordinates are formed out of polynomials in $x$ and thus
  constitute a vector space of polynomials in $x$. This vector space
  can be equipped either with the standard pointwise multiplication or
  with a deformed multiplication which is comprised within the above
  defined star product. We shall adequately denote the corresponding
  algebras with ${\mathcal{A}}$ and ${\mathcal{A}}_{a},$ respectively.
 This means that the vector space of fields in commutative coordinates,
  equipped with usual pointwise multiplication, forms a symmetric
  algebra ${\mathcal{A}},$ generated by commuting coordinates $x_{\mu},$
 while the same structure, which has a star product as a multiplication
  operation, constitutes a deformed algebra ${\mathcal{A}}_{a}.$ 
While the algebra ${\mathcal{A}}$ is undeformed and can be considered
as a trivial example of an universal enveloping algebra of functions
in commutative coordinates, the later one, that is algebra ${\mathcal{A}}_{a},$
is deformed one, with deformation being encoded within a star product.
Commutative fields $\phi(x)$ that are formed from polynomials in
commutative coordinates thus can either be considered as if they
belong to an universal enveloping algebra ${\mathcal{A}},$ generated by commuting
coordinates $x_{\mu},$ or to a deformed algebra ${\mathcal{A}}_{a},$
where a deformation is encoded within the star product.
 In this setting, the unit element
$1, $  having the properties,
\begin{eqnarray}  \label{2.22}
\phi(x) \triangleright 1 &  = ~ \phi(x), \qquad \hat{\phi}(\x)
\triangleright 1 ~ = ~ \phi(x), \\
\p_{\mu} \triangleright 1 &  = ~0, \qquad  M_{\mu\nu} \triangleright 1 ~ = ~ 0,  \label{2.23}
\end{eqnarray}
 can be thought of as the unit element in the universal enveloping
 algebra ${\mathcal{A}}$ (or more precisely, in a completion of the symmetric algebra ${\mathcal{A}}$)
understood as a module over the  deformed Weyl
 algebra, which is generated by
  $\x_{\mu}$ and $\p_{\mu}, ~~ \mu = 0,1,...,n-1,$ and allows for 
  infinite series in $\p_{\mu}$. It is also understood that
NC coordinates  $\x,$ appearing in (\ref{2.22}),  refer to particular realization (\ref{2.10}),
i.e. they are assumed to be represented by (\ref{2.10}). 

\section{Hopf algebra structure}

A deformation of Minkowski spacetime made in accordance with  commutation
  relations (\ref{2.1}) ($\kappa$-deformation) implies some important questions that have to
  be addressed. One of them is the question on the real nature of symmetry describing 
  $\kappa$-deformed Minkowski space, particularly the question whether the 
 Poincar\'{e} symmetry is still a relevant symmetry for  theories
built on such  $\kappa$-deformed spaces or instead Poincar\'{e} symmetry itself is also affected by deformation.
In analysing these issues, we are naturally led to the conclusion that
   symmetry underlying $\kappa$-deformed Minkowski space is deformed too.
This $\kappa$-deformed Poincar\'{e} symmetry can most conveniently be described in
  terms of quantum Hopf algebra \cite{majid1}. Besides an algebraic part, which we take
  by our choice as undeformed and thus 
  described by relations (\ref{2.2}),(\ref{2.4}) and (\ref{2.5}), the full
  description of
$\kappa$-deformed Poincar\'{e} symmetry also requires an information
on the action of Poincar\'{e} generators on the enveloping algebra
  ${\hat{\mathcal{A}}}_{a}$, i.e. a type of information which is encoded within the coalgebraic
  part of the corresponding Hopf algebra. As we shall see, the action
  of Poincar\'{e} generators  on 
   deformed Minkowski spacetime and the algebra ${\hat{\mathcal{A}}}_{a}$ is
  deformed. Since deformed action of Poincar\'{e} generators leads
  to a deformed coproduct structure and deformed
  Leibniz rules, we have as a conclusion
  that the  Hopf algebra, describing $\kappa$-deformation of Minkowski
  space,  is characterized by a
  deformed coalgebraic sector, while simultaneously having an undeformed
  algebraic part, as discussed above. In this way, the whole 
deformation is contained within the coalgebraic sector alone.

 We now present the basic ingredients of the mathematical structure in
 question as follows. Denoting $\kappa$-Poincar\'{e} algebra by ${\bf{g}},$
then as a Lie algebra, it has a unique universal enveloping algebra ${\mathcal{U}}_{a}({\bf{g}})$
which maintains a Lie algebra structure preserved. The algebra ${\mathcal{U}}_{a}({\bf{g}})$
becomes a Hopf algebra if it is endowed with a coalgebra structure,
 i.e. coproduct
$\triangle : {\mathcal{U}}_{a}({\bf{g}}) \longrightarrow
 {\mathcal{U}}_{a}({\bf{g}}) \otimes {\mathcal{U}}_{a}({\bf{g}}), $
counit $\epsilon : {\mathcal{U}}_{a}({\bf{g}}) \longrightarrow {\bf{C}} $
and antipode $S : {\mathcal{U}}_{a}({\bf{g}}) \longrightarrow {\mathcal{U}}_{a}({\bf{g}}). $
In this case the algebra ${\hat{\mathcal{A}}}_{a}$ can be understood as a
 module algebra for ${\mathcal{U}}_{a}({\bf{g}})$ since the elements of ${\mathcal{U}}_{a}({\bf{g}})$
act on it by Hopf action.

The algebraic part of the $\kappa$-Poincar\'{e} Hopf algebra ${\mathcal{U}}_{a}({\bf{g}})$ is given in relations
(\ref{2.2}),(\ref{2.4}) and (\ref{2.5}). The coalgebraic part, which
 includes coproducts
  for translation ($P_{\mu}= -i\p_{\mu}$), rotation and boost generators
 \cite{KresicJuric:2007nh},\cite{Meljanac:2007xb}, is given by the
 following relations,
\begin{eqnarray} \label{coproductmomentum}
\triangle \p_{\mu} &=& \p_{\mu}\otimes Z^{-1}+\mathbf{1}\otimes
\p_{\mu}+ia_{\mu} (\p_{\lambda} Z)\otimes
\p^{\lambda}-\frac{ia_{\mu}}{2} \square\, Z\otimes ia\p, \\
\triangle M_{\mu\nu} &=& M_{\mu\nu}\otimes
\mathbf{1}+\mathbf{1}\otimes M_{\mu\nu} \nonumber \\
&+& ia_{\mu}\left(\p^{\lambda}-\frac{ia^{\lambda}}{2}\square\right)\,
Z\otimes
M_{\lambda\nu}-ia_{\nu}\left(\p^{\lambda}-\frac{ia^{\lambda}}{2}\square\right)\,
Z\otimes M_{\lambda\mu}. \label{coproductangmomentum}
\end{eqnarray}
In the above expressions $Z$ is the shift operator, Eq.(\ref{16}), whose
 copruduct is simply given by
\begin{equation}
\triangle Z = Z\otimes Z.
\end{equation}
The operator $\square$ is a deformed d'Alambertian operator \cite{Dimitrijevic:2003wv},\cite{Meljanac:2006ui},\cite{KresicJuric:2007nh},
\begin{equation} \label{3.7}
  \square = \frac{2}{a^2}(1- \sqrt{1- a^2 \p^2}), 
\end{equation}
with the property $\square \rightarrow \p^2 \;$ as $a \rightarrow 0$.
We now turn to antipodes for the generators of $\kappa$-Poincar\'{e} algebra.
The antipode for translation generators $P_{\mu}$ can be
written in a compact way as
\begin{equation} \label{3.8}
 S(P_{\mu}) = \left( -P_{\mu} - a_{\mu}P^2 +
 \frac{1}{2}a_{\mu}(aP)\square (P) \right) Z(P),
\end{equation}
where, in accordance with (\ref{16}) and (\ref{3.7}),
\begin{equation} \label{3.8a}
   Z^{-1} (P) = aP + \sqrt{1 + a^2 P^{2}}, \qquad
 \square (P) = \frac{2}{a^2}(1- \sqrt{1+ a^2 P^2}).   
\end{equation}
On the other hand, antipode for Lorentz generators has the form
\begin{equation} \label{3.8b}
 S(M_{\mu \nu}) = -M_{\mu \nu}
   + a_{\nu} \left ( P_{\alpha} - \frac{a_{\alpha}}{2} \square (P)
   \right ) M_{\alpha \mu}
  - a_{\mu} \left ( P_{\alpha} - \frac{a_{\alpha}}{2} \square (P)
   \right ) M_{\alpha \nu}.
\end{equation}
Finally counits of the generators of $\kappa$-Poincar\'{e} algebra remain trivial.
One can also check that the coassociativity condition for the coproduct is fulfilled,
\begin{equation} \label{3.9}
 (id \otimes \triangle) \triangle = (\triangle \otimes id) \triangle,
\end{equation}
with $id : {\mathcal{U}}_{a}({\bf{g}}) \longrightarrow {\mathcal{U}}_{a}({\bf{g}}) $
being an identity operator.

The antipode for translation generators has some usefull properties:
\begin{eqnarray} \label{3.10a}
i) &&  S(P_{\mu})S(P^{\mu}) = P_{\mu}P^{\mu}, \quad  \mbox{or simply}  \quad
  {(S(P))}^2 = P^2, \\
ii) &&  S(S(P_{\mu})) = P_{\mu}, \label{3.10b} \\
iii) &&  Z^{-1}(S(P)) = Z(P),  \label{3.10c} \\
iv) &&  \square (S(P)) = \square (P).  \label{3.10d}
\end{eqnarray}
We shall make an extensive use of the above notions and their
 properties in considerations that will follow shortly and
 in building the field
  theory on NC spacetime with noncommutative structure (\ref{2.1}).

\section{Free scalar field theory in nonhermitian realization}

The realization (\ref{2.10}), as it stands is nonhermitian one, i.e.
for $\hat{x}_{\mu}$ in (\ref{2.10}) the relation
${\hat{x}_{\mu}}^{\dagger} = \hat{x}_{\mu} $ does not hold! However,
this realization has some convenient and usefull properties \cite{Meljanac:2007xb},\cite{KresicJuric:2007nh}.
It has been shown in Refs.\cite{liealgebradef},\cite{Meljanac:2009fy}
that for $\hat{x}_{\mu},$ given in (\ref{2.10}), one can write the
following identities:
\begin{equation} \label{4.1}
 e^{iP\x} \triangleright 1 ~ = ~ e^{i{\bf{K}}_{\mu}(P)x^{\mu}}
\end{equation}
and
\begin{equation} \label{4.2}
 e^{iP\x}  \triangleright e^{iQx} ~ = ~ e^{i{\bf{P}}_{\mu}(P,Q)x^{\mu}},
\end{equation}
where the unit element $1$ is defined in (\ref{2.19}),
 $P\x = P^{\alpha} {\x}_{\alpha} $ and 
 quantities  ${\bf{K}}_{\mu}(P),$ ${\bf{P}}_{\mu}(P,Q)$ have the form
\begin{eqnarray} \label{4.2a}
 {\bf{P}}_{\mu}(P,Q) & = & Q_{\mu} + \left( P_{\mu} Z^{-1}(Q) -a_{\mu}
 (PQ)  \right) \frac{\sinh(aP)}{aP} \nonumber \\
        &+& \bigg[ \left(P_{\mu}(aP) - a_{\mu}P^2 \right) Z^{-1}(Q)
     + a_{\mu}(aP) (PQ) \bigg] \frac{\cosh (aP)-1}{{(aP)}^2}, \\
  {\bf{K}}_{\mu}(P) = {\bf{P}}_{\mu}(P,0) &=&  P_{\mu} \frac{e^{aP}-1}{aP} -
  a_{\mu} P^2 \frac{\cosh (aP) -1}{{(aP)}^2}. \label{4.2b} 
\end{eqnarray}
As before (see Eq.(\ref{3.8a})), the quantity $Z^{-1} (Q)$ can be
expressed as
\begin{equation} \label{4.3}
  Z^{-1} (Q) ~  = ~ aQ + \sqrt{1 + a^2 Q^2} ~ = ~ e^{a {\bf{K}}^{-1}(Q)},
\end{equation}
where
\begin{equation} \label{4.4}
  {\bf{K}}^{-1}_{\mu}(P) = \bigg[ P_{\mu} - \frac{a_{\mu}}{2} \square(P)
       \bigg] \frac{\ln \left( Z^{-1} (P) \right)}{ Z^{-1} (P) -1}
\end{equation}
is the inverse transformation of (\ref{4.2b}).
It is also understood that quantities like $(PQ)$ have the meaning of the scalar
product between $n$-momenta $P$ and $Q$ in a Minkowski space with signature
 $~ \eta_{\mu\nu} = diag (-1,1,\cdot \cdot \cdot, 1)$ and that
 $n$-momenta $P$ and $Q$ allow the identification with the operator
$-i\p ~$ (e.g. $Q=-i\p$), as discussed at the beginning.
 
According to relation (\ref{4.1}) and the definition of the star product introduced in (\ref{2.21}),
we can write for the star product between two plane waves
\begin{equation} \label{4.11}
  e^{iPx}~ \star ~ e^{iQx} ~ = ~ e^{i {\bf{K}}^{-1}(P) \x} \triangleright e^{iQx}
 ~ = ~ e^{i {\mathcal{D}}_{\mu} (P,Q) x^{\mu}}, 
\end{equation}
where, in accordance with (\ref{4.2}),
\begin{equation} \label{4.12}
   {\mathcal{D}}_{\mu} (P,Q) ~ = ~ {\bf{P}}_{\mu} ({\bf{K}}^{-1}(P),Q),
\end{equation}
with ${\bf{K}}^{-1}(P)$ being given in (\ref{4.4}).
It can readily be shown that the shift operator
 $~Z^{-1}(P)~$ and the d'Alambertian operator $~\square(P) ~$ can be expressed in terms
of the quantity ${\bf{K}}^{-1}(P)$ as
\begin{equation} \label{4.13}
  Z^{-1} (P) ~ \equiv ~ aP + \sqrt{1 + a^2 P^2} ~ =~ e^{a {\bf{K}}^{-1}(P)},
\end{equation}
\begin{equation} \label{4.14}
   \square(P) ~ \equiv ~ \frac{2}{a^2}\bigg[ 1- \sqrt{1 + a^2 P^2}
   \bigg] ~ = ~ 2  \frac{1-
 \cosh \left( a {\bf{K}}^{-1}(P) \right) }{{\left(a {\bf{K}}^{-1}(P)
   \right)}^2} {({\bf{K}}^{-1}(P))}^2.
\end{equation}
We can make use of relations (\ref{4.13}) and (\ref{4.14}) to find
 the function ${\mathcal{D}}(P,Q) $ in (\ref{4.12}), which determines 
   the momentum addition rule, $~ {\mathcal{D}}(P,Q) = P \oplus Q,~$ in $\kappa$-deformed
 Minkowski space (\ref{2.1}). This generalized rule for addition of
 momenta turns out \cite{KresicJuric:2007nh},\cite{Meljanac:2007xb} to have the form 
\begin{equation} \label{4.16}
 {\mathcal{D}}_{\mu}(P,Q) = ~(P \oplus Q)_{\mu} ~ = ~ P_{\mu} Z^{-1} (Q)  + Q_{\mu} - a_{\mu}(PQ)Z(P) 
    +  \frac{1}{2} a_{\mu}(aQ) \square (P) Z(P).
\end{equation} 

A comparison of Eq.(\ref{4.16}) against expression (\ref{coproductmomentum}),
 which gives the coproduct for translation generators, $\triangle \p_{\mu}$,
 reveals that it is possible to make an identification
\begin{equation} \label{4.12a}
  i{\mathcal{D}}_{\mu} (-i\p \otimes 1, 1 \otimes (-i\p)) = \triangle
  \p_{\mu}, 
\end{equation} 
showing that function $~ {\mathcal{D}} (P,Q) = P \oplus Q, ~$ besides giving the
rule for adding momenta, also comprises a deformed
Leibniz rule and the corresponding coproduct for translation
generators of $\kappa$-Poincar\'{e} algebra.
This result, i.e. Eq.(\ref{4.12a}), is 
 completely consistent
 with the general definition of the star product, 
\begin{equation} \label{4.12b}
 f~ \star ~ g = m_{\star} (f \otimes g)
\end{equation} 
and with the expression describing mutual relationship between star
product and coproduct,
\begin{equation} \label{4.12c}
 \p_{\mu} (f \star g) = m_{\star} (\triangle \p_{\mu} (f \otimes g)),
\end{equation} 
which is the relation that immediately follows from (\ref{4.12b}) after taking a derivative on both sides in
  (\ref{4.12b}). In the above two expressions
 $m_{\star} : {\mathcal{A}}_{a} \otimes {\mathcal{A}}_{a} \longrightarrow {\mathcal{A}}_{a} $ denotes a deformed multiplication in the
algebra of commutative and smooth functions. 
After applying (\ref{4.12c}) to (\ref{4.11}), Eq.(\ref{4.12a}) follows immediately,
which, as we have seen, is the result expected on the ground of comparison between Eqs.(\ref{4.16})
 and (\ref{coproductmomentum}). This shows an internal consistency of the entire setting
   we are working in,
since different approaches bring about the same conclusion.

With the known coproduct, it is a straightforward procedure
 \cite{Meljanac:2006ui},\cite{Meljanac:2007xb}
to find a star product between two arbitrary
 elements $f$ and $g$ in the algebra ${\mathcal{A}}_{a}$, generalizing in this way relation (\ref{4.11}) that
holds for plane waves. Thus, the star product, describing 
noncommutative features of the spacetime with commutation relations
  (\ref{2.1}), has the form
\begin{equation} \label{4.17}
(f \; \star \; g)(x)  =   \lim_{\substack{u \rightarrow x }}
 m \left ( e^{x^{\alpha} ( \triangle - {\triangle}_{0}) {\partial}_{\alpha} }
    f(u) \otimes g(u) \right ), 
\end{equation}
where $ \; {\triangle}_{0}{\partial}_{\mu} =
 {\partial}_{\mu} \otimes 1 + 1 \otimes {\partial}_{\mu}, \; $
$ \: \triangle {\partial}_{\mu} \; $ is given in (\ref{coproductmomentum})
 and $ \; m : {\mathcal{A}} \otimes {\mathcal{A}} \longrightarrow {\mathcal{A}} \; $
 is the undeformed multiplication map in the algebra ${\mathcal{A}}$,
 namely, $ \; m (f(x) \otimes g(x)) = f(x) g(x). $
The expression (\ref{4.17}) gives a general form for the star product on
$\kappa$-Minkowski spacetime (\ref{2.1}).
The star product (\ref{4.17}) with the coproduct
 (\ref{coproductmomentum}) acts
 as a deformed
 multiplication map
 in the algebra ${\mathcal{A}}_{a}$ of smooth, commutative functions. It
 also reflects the noncommutative
 nature of $\kappa$-Minkowski spacetime and 
due to coassociativity of the coproduct, Eq.(\ref{3.9}), is associative.
The opposite is also true, an associativity of star
 product (\ref{4.17}) implies coassociativity of the coproduct, Eq.(\ref{3.9}).
 
 In what follows we shall use the results obtained so far in order to
 build a massive, generally complex, scalar field theory on $\kappa$-deformed Minkowski
 space and to relate it to a corresponding field theory on undeformed
 Minkowski space. A proper construction of field theory, besides using
 a star product, i.e. deformed multiplication  between the elements of
 algebra ${\mathcal{A}}_{a},$ instead of usual pointwise
 multiplication between these elements, also requires an introduction of the
 adjoint (or hermitian conjugation) operation $\dagger,$ as an obverse to common complex
 conjugation operation used in standard field theory.
 However, at this
 place it should be emphasized that the adjoint operation $\dagger$ will have
 different actions when applied to the elements of deformed algebras
 ${\mathcal{A}}_{a}$ and ${\hat{\mathcal{A}}}_{a},$ respectively
and since we want to establish a full
 correspondence between these two deformed algebras
(through the map $\Omega$), we also have to know the respective actions of
 the adjoint operation $\dagger,$ performed on the corresponding algebras ${\mathcal{A}}_{a}$ and ${\hat{\mathcal{A}}}_{a}$. 
Turned in other way, this means that the adjoint operation $\dagger$
 induces two different operations $\dagger$, one exhibited when applied to the
 elements of algebra ${\mathcal{A}}_{a}$ of commutative coordinates and
 the other when applied to the elements of algebra ${\hat{\mathcal{A}}}_{a}$
 of noncommutative coordinates. In what follows we shall take the same
 symbol $\dagger$ for both of these operations, but it should be always kept
 in mind that they are mutually different
\footnote{The adjoint operation $\dagger$ which acts on the elements of ${\hat{\mathcal{A}}}_{a}$
   is a standard one and is characterized by the property
  ${P_{\mu}}^{\dagger} = {(-i {\hat{{\p}}}_{\mu})}^{\dagger} \equiv {(-i {\p}_{\mu})}^{\dagger} = P_{\mu}, $
   while the adjoint operation $\dagger$ which acts on the elements of ${\mathcal{A}}_{a}$
 is characterized by the property ${P_{\mu}}^{\dagger} = {(-i {\p}_{\mu})}^{\dagger} = -S(P_{\mu}), $
 with $S(P_{\mu})$ beeing the antipode (\ref{3.8}). We recall
 that  ${\mathcal{A}}_{a}$ is the algebra of commutative functions
 whose multiplication is being given by the star product.
Thus, there should be no problem to distinguish between these two adjoint operations since
 each of them acts only on ${\hat{\mathcal{A}}}_{a}$ or ${\mathcal{A}}_{a},$
 respectively. Thus, whenever we have ${\x}^{\dagger}$ or
 ${\hat{\phi}}^{\dagger} ({\x}^{\dagger}),$ it is understood that
 $\dagger$ means the adjoint operation defined on ${\hat{\mathcal{A}}}_{a},$ while
 whenever we have $x^{\dagger}$ or ${\phi}^{\dagger} (x),$ it is understood that
 $\dagger$ stands for the adjoint operation defined on ${\mathcal{A}}_{a}.$}. 

In the case that one deals with realization of $\x$ which 
is not hermitian, as is the case with realization (\ref{2.10}),
the construction of scalar field theory
 cannot be done in a way consistent with prescriptions (\ref{2.20})
and (\ref{2.21}), because in these circumstances noncommutative plane
waves will not be unitary operators and thus
 it is not clear at all how should adjoint operator of $e^{iP\x}$ look like
and what plane wave in commutative coordinates this adjoint operator would correspond
to through the map $\Omega,$ Eq.(\ref{2.20}). In the next section we shall demonstrate how these type of problems
can be overcome. In particular, we shall see how we can achieve that for every adjoint of
the usual commutative plane wave, a corresponding adjoint of
noncommutative plane wave can be uniquely defined in a way that is
consistent with prescriptions (\ref{2.20})
and (\ref{2.21}). 

We take the adjoint of the standard
 plane wave to be defined as \cite{Freidel:2006gc},\cite{Meljanac:2007xb}
\begin{equation} \label{4.18}
 {\left(e^{iPx} \right)}^{\dagger} = e^{iS(P)x},
\end{equation}
where $S(P)$ is antipode (\ref{3.8}) and $\dagger$ means the adjoint
   operation defined on ${\mathcal{A}}_{a}$ (see the footnote 1). In the next section this
relation will be justified on the more fundamental ground, but for the
moment we just take over it without going into the details.
The standard, commutative, generally complex scalar field is assumed to have a Fourier expansion
\begin{equation} \label{4.19}
 \phi (x) = \int d^n P ~ \tilde{\phi} (P) ~ e^{iPx}.
\end{equation}
Now we can write the action for non-interacting complex massive scalar field as
\begin{align} \label{4.20}
 S[\phi] & = \int d^n x ~ {\mathcal{L}}(\phi, \p_{\mu}\phi) \nonumber \\
 & = \frac{1}{2}\int d^n x ~ {(\p_{\mu}\phi)}^{\dagger} \star (\p^{\mu}\phi) +
 \frac{m^2}{2}\int d^n x ~ {\phi }^{\dagger} \star \phi. 
\end{align}
In order to find its relation with the corresponding action
on undeformed Minkowski spacetime, we shall consider an integral
expression of the form
\begin{align} \label{4.21}
 \int d^n x ~ {\psi }^{\dagger} \star \phi, 
\end{align}
and rewrite it in terms of pointwise multiplication between arbitrary
elements $\psi$  and $\phi$
in the algebra ${\mathcal{A}}$. This will result in a very usefull
integral mathematical identity that will have a crucial role in our analysis.
According to Eqs.(\ref{4.18}) and (\ref{4.19}) we have a Fourier
expansion for the adjoint of the field $\phi,$
\begin{equation} \label{4.22}
 {\phi}^{\dagger} (x) = \int d^n P ~ {\tilde{\phi}}^{\ast} (P) ~ e^{iS(P)x},
\end{equation}
where $*$ denotes the standard complex conjugation.
Consequently, by anticipating deformed multiplication between
plane waves, Eq.(\ref{4.11}), it follows
\begin{eqnarray} \label{4.23}
 \int d^n x ~ {\psi }^{\dagger} \star \phi & = &
 \int d^n x \int d^n P \int d^n Q ~ {\tilde{\psi}}^{\ast} (P)
 {\tilde{\phi}} (Q) ~ e^{iS(P)x} ~ \star ~ e^{iQx} \nonumber \\
  & = & \int d^n P \int d^n Q ~ {\tilde{\psi}}^{\ast} (P)
 {\tilde{\phi}} (Q) \int d^n x ~ e^{i{\mathcal{D}}(S(P),Q)x}   \nonumber \\
 & = & \int \int d^n P d^n Q ~ {\tilde{\psi}}^{\ast} (P)
 {\tilde{\phi}} (Q) {(2 \pi)}^n {\delta}^{(n)} \left( {\mathcal{D}}(S(P),Q)\right),
\end{eqnarray}
where the quantity in the argument of ${\delta}^{(n)}$-function can be deduced
from (\ref{3.8}),(\ref{4.16}) and from the properties (\ref{3.10a})-(\ref{3.10d}),
\begin{eqnarray} \label{4.24}
 {\mathcal{D}}_{\mu}(S(P),Q) & = & ~ \left( -P_{\mu} - a_{\mu}P^2 +
 \frac{1}{2}a_{\mu}(aP)\square (P) \right) Z(P) Z^{-1} (Q)  + Q_{\mu}
 + a_{\mu}(PQ) \nonumber \\
    & + &  a_{\mu}(aQ) P^2 - \frac{1}{2} a_{\mu}(aQ)(aP) \square (P)
 +  \frac{1}{2} a_{\mu}(aQ) \square (P) Z^{-1}(P).
\end{eqnarray} 
$Z^{-1}(P)$ and $\square (P)$ above are given by Eqs.(\ref{4.13}) and
(\ref{4.14}), respectively. To calculate ${\delta}^{(n)}$-function in
(\ref{4.23}), we use the identity
\begin{equation} \label{4.25}
 {\delta}^{(n)} \left( F(P,Q)\right) = {\sum}_{i} \frac{{\delta}^{(n)}
 (Q-Q_i )}{{\left|  \det \left( \frac{\p F_{\mu}(P,Q)}{\p Q_{\nu}} \right) \right|}_{Q=Q_i}},
\end{equation}
where the expression in denominator is $n\times n$ Jacobi determinant
of the transformation $~ Q \longmapsto F(P,Q),~$ with the reminder
that while $Q$ is treated as an independent variable, $P$ is assumed
to be a parameter.
On the other hand $Q_i$ are zeros of the generic function $F,$ which, for the need of our calculation, we specialize
in this case to $~F(P,Q) \equiv {\mathcal{D}}(S(P),Q). ~$ To proceed, we need
matrix entries $\frac{\p {\mathcal{D}}_{\mu}(S(P),Q)}{\p Q^{\lambda }}$ of Jacobian, which are given by
\begin{eqnarray} \label{4.26}
 \frac{\p {\mathcal{D}}_{\mu}(S(P),Q)}{\p Q^{\lambda }} & = & ~ \left( -P_{\mu} - a_{\mu}P^2 +
 \frac{1}{2}a_{\mu}(aP)\square (P) \right) Z(P) \left(a_{\lambda} +
 \frac{a^2}{\sqrt{1+a^2 Q^2}} Q_{\lambda} \right) \nonumber \\
 & + & {\eta}_{\mu \lambda} + a_{\mu}P_{\lambda} 
    +  a_{\mu}a_{\lambda} P^2 + \frac{1}{2} a_{\mu}a_{\lambda} \square (P)
   \sqrt{1 + a^2 P^2}.
\end{eqnarray} 
For simplicity and just for the sake of calculation of the determinant,
 we take the  four-vector of the deformation parameter $a$ to be
colinear with the time direction, $~ a = (a_0 , 0,...,0), ~$ and write the resulting Jacobian in
corresponding Lorentz frame, 
\begin{equation} \label{Jacobian}
  \det {\left( \frac{\p {\mathcal{D}}_{\mu}(S(P),Q)}{\p Q^{\lambda }} \right)}_{Q=P} =
\left| \begin{array}{ccccc}
  -c & bZ(P)a_{0}^2 c P_1 + a_0 P_1  & 
   \cdot \cdot \cdot & bZ(P)a_{0}^2 c P_{n-2}
 + a_0 P_{n-2}  & ~  bZ(P)a_{0}^2 c P_{n-1} + a_0 P_{n-1} \\
 -a_{0}cP_1  & 1+ Z(P) a_{0}^2 c P_{1}^2 &  \cdot \cdot \cdot
 & Z(P) a_{0}^2 c P_{1} P_{n-2} & Z(P) a_{0}^2 c P_{1} P_{n-1} \\ 
 \cdot & \cdot &   \cdot \cdot \cdot & \cdot & \cdot \\
\cdot & \cdot &  \cdot \cdot \cdot & \cdot & \cdot \\
 -a_{0}cP_{n-3}  &  Z(P) a_{0}^2 c P_{n-3} P_{1} &  \cdot \cdot \cdot 
 & Z(P) a_{0}^2 c P_{n-3} P_{n-2} & Z(P) a_{0}^2 c P_{n-3} P_{n-1} \\
  -a_{0}cP_{n-2} & Z(P) a_{0}^2 c P_{n-2} P_{1} & \cdot \cdot \cdot
 & 1+ Z(P) a_{0}^2 c P_{n-2}^2 & Z(P) a_{0}^2 c P_{n-2} P_{n-1} \\
  -a_{0}cP_{n-1}  & Z(P) a_{0}^2 c P_{n-1} P_{1} & \cdot \cdot \cdot 
 & Z(P) a_{0}^2 c P_{n-1} P_{n-2} & 1+ Z(P) a_{0}^2 c P_{n-1}^2 \\
\end{array} \right|
\end{equation}
where the  quantities $ c$ and $b$ are defined as
\begin{equation} \label{4.27}
  c = \frac{1}{\sqrt{1- a_{0}^2 P^2}},   \qquad    b = a_{0} P^2 + P_{0} \sqrt{1- a_{0}^2 P^2}.
\end{equation}
The Jacobian (\ref{Jacobian}) can be calculated to give
\begin{eqnarray} \label{4.28}
 \det {\left( \frac{\p {\mathcal{D}}_{\mu}(S(P),Q)}{\p Q^{\lambda }} \right)}_{Q = P} 
   & = & ~ - \frac{1}{\sqrt{1- a_{0}^2 P^2}},  \qquad   \mbox{and thus}
    \nonumber \\
 \det {\left( \frac{\p {\mathcal{D}}_{\mu}(S(P),Q)}{\p Q_{\lambda }} \right)}_{Q = P}
& = & - \det {\left( \frac{\p {\mathcal{D}}_{\mu}(S(P),Q)}{\p Q^{\lambda }} \right)}_{Q = P}
 = ~ \frac{1}{\sqrt{1+ a^2 P^2}}.
\end{eqnarray}
Due to the fact that all quantities are written in a covariant
fashion, we can do the same with the final result for the determinant (\ref{Jacobian})
and rewrite it also in a covariant form, giving rise to the last line in
(\ref{4.28}). In fact, all we had to do was to rewrite $a_0^2$ in a covariant
form, $a_0^2 = - a^2,$ respecting the signature we work with.
 As a consequence of the
definition of the antipode, $~{\mathcal{D}}(S(P), P) = {\mathcal{D}}(P, S(P)) = 0, \; $ there is
only one zero for the expression in (\ref{4.24}). This zero is the trivial
one, $~ Q= P, ~$ and was used in calculating the Jacobian (\ref{Jacobian}).
 Now we have
\begin{equation} \label{4.29}
 {\delta}^{(n)} \left( {\mathcal{D}}(S(P),Q)\right) ~ = ~
 \sqrt{1+ a^2 P^2} ~ {\delta}^{(n)} (Q-P ).
\end{equation}
By plugging this into (\ref{4.23}) and using $P_{\mu} = -i{\p}_{\mu}, \; $ 
as well as the integral representation of ${\delta}^{(n)}$-function,
\begin{equation} \label{4.30}
  {\delta}^{(n)} (Q-P ) ~ = ~ \frac{1}{{(2\pi)}^n} \int d^n x ~ e^{i(Q-P)x},
\end{equation}
we get the following integral identity
\begin{equation} \label{4.31}
 \int d^n x ~ {\psi }^{\dagger} \star \phi  ~ = ~
 \int d^n x ~ {\psi }^{\ast} (x) \sqrt{1- a^2 {\p}^2} ~ \phi (x), 
\end{equation}
where the symbol $*$ denotes the usual complex conjugation in the
undeformed algebra ${\mathcal{A}}$.
When the identity (\ref{4.31}) is applied to the action (\ref{4.20}),
we get
\begin{equation} \label{4.32}
 S[\phi] = \frac{1}{2} \int d^n x ~
  \left[ {({\p}_{\mu} \phi )}^{\ast} (x) \sqrt{1- a^2 {\p}^2} ~
  ({\p}^{\mu}\phi ) (x)
 + m^2 {\phi }^{\ast} (x) \sqrt{1- a^2 {\p}^2} ~ \phi (x) \right],
\end{equation}
which is the action that describes a nonlocal, relativistically
invariant free scalar field theory on undeformed Minkowski spacetime
\cite{Freidel:2006gc},\cite{Meljanac:2007xb}. The action (\ref{4.32})
is also translationally invariant, since the star product
(\ref{4.11}), i.e. (\ref{4.17}) is translationally invariant. This can
indeed be verified by utilising that $\kappa$-Minkowski space is
invariant \cite{Kosinski:1999dw},\cite{Daszkiewicz:2007eq} under translations 
$ \x_{\mu} \longrightarrow \x_{\mu}  + \hat{y}_{\mu} $ if
\begin{equation} \label{4.33}
 [ \hat{y}_{\mu}, \hat{y}_{\nu}] = i(a_\mu \hat{y}_{\nu} - a_\nu \hat{y}_{\mu}),
 \qquad [ \x_{\mu}, \hat{y}_{\nu}] = 0.
\end{equation}
Another point that we anticipate is that in terms of realizations, on which our approach is
based, we can write
\begin{equation} \label{4.34}
 \x_{\mu} =x^\alpha \Phi_{\alpha\mu}(\p_x), \qquad  {\hat{y}}_{\mu} =y^\alpha \Phi_{\alpha\mu}(\p_y),
\end{equation}
where $\p_x$ and $\p_y$ are derivatives with respect to different sets
of commutative coordinates $x$ and $y,$ respectively. 
Now, by taking into account relations
\begin{equation} \label{4.35}
   e^{i {\bf{K}}^{-1}(P) \x} \triangleright 1
 ~ = ~ e^{i Px}, \qquad  e^{i {\bf{K}}^{-1}(Q) \x} \triangleright 1
 ~ = ~ e^{i Qx},
\end{equation}
and by making use of (\ref{4.33}), we have
\begin{equation} \label{4.36}
   e^{i {\bf{K}}^{-1}(P) (\x + \hat{y})} \triangleright 1
 ~ = ~  e^{i {\bf{K}}^{-1}(P) \x} e^{i {\bf{K}}^{-1}(P) \hat{y}} \triangleright 1
~ = ~ e^{i Py} e^{i {\bf{K}}^{-1}(P) \x} \triangleright 1     ~ = ~ e^{i P(x+y)},
\end{equation}
and similarly
\begin{equation} \label{4.37}
   e^{i {\bf{K}}^{-1}(Q) (\x + \hat{y})} \triangleright 1
 ~ = ~ e^{i Q(x+y)}.
\end{equation}
Consequently we can write
\begin{eqnarray} \label{4.38}
 e^{i P(x+y)} \star e^{i Q(x+y)} ~ & = & ~  e^{i {\bf{K}}^{-1}(P) (\x
    + \hat{y})} e^{i {\bf{K}}^{-1}(Q) (\x + \hat{y})} \triangleright 1    \nonumber \\
  ~ & = & ~ e^{i{\mathcal{D}}(P,Q)x} e^{i{\mathcal{D}}(P,Q)x} ~ = ~
 e^{i{\mathcal{D}}(P,Q) (x + y)},
\end{eqnarray}
showing that the star product in nonhermitian realization is translationally invariant
and so is the action (\ref{4.32}).

\section{Free scalar field theory in hermitian realization}

The aim of this section is to finally resolve the open problems that are 
 addressed, but not in any case answered in the previous section. 
It specifically applies to a problem of proper definition of the
 adjoint operation $\dagger,$ which would be consistent with the map $\Omega$
 between  algebra ${\mathcal{A}}_{a}$ of commutative functions and
 enveloping algebra ${\hat{\mathcal{A}}}_{a}$ of noncommutative functions
   and with the definition (\ref{2.21}) of the star product, as well
 as with the whole construction of the field
theoretic action.
In the previous section we have worked with realization (\ref{2.10})
which is not hermitian. However, it can be made hermitian by forming the combination
\begin{equation} \label{5.1}
 {\x}^{h}_{\mu} = \frac{1}{2} ({\x}_{\mu} + {\x}^{\dagger}_{\mu}),
\end{equation}
where $\dagger$ means the adjoint
   operation defined on ${\hat{\mathcal{A}}}_{a}$
    (this operation is induced by the standard hermitian conjugation in
   ${\mathcal{A}}$, i.e. ${({\p}_{\mu})}^{\dagger} = -{\p}_{\mu}, ~
   {x_{\mu}}^{\dagger} = x_{\mu}, $ see the footnote 1).
This leads to realization
\begin{eqnarray} \label{5.2}
 {\x}^{h}_{\mu} & = & {\x}_{\mu} - \frac{1}{2} \frac{a^2
 {\p}_{\mu}}{\sqrt{1-a^2 {\p}^2}} \nonumber \\
  & = & x_{\mu}(-ia_{\alpha}{\p}^{\alpha}+\sqrt{1-a^2{\p}_{\alpha}{\p}^{\alpha}})+i(ax){\p}_{\mu}
 - \frac{1}{2} \frac{a^2 {\p}_{\mu}}{\sqrt{1-a^2 {\p}^2}},
\end{eqnarray}
which is  hermitian by construction, ${({\hat{x}}^{h}_{\mu})}^{\dagger} = {\hat{x}}^{h}_{\mu}, $
and satisfies Eq.(\ref{2.1}), i.e.
\begin{equation} \label{5.3}
 [{\x}^{h}_{\mu},{\x}^{h}_{\nu}] = i(a_{\mu}{\x}^{h}_{\nu}-a_{\nu}{\x}^{h}_{\mu}), 
\end{equation}
as well as other commutation relations, Eqs.(\ref{2.2})-(\ref{2.6}).
Thus, all commutation relations and particularly (\ref{2.3}) and (\ref{2.6})
remain unchanged, when expressed in terms of ${\x}^{h}_{\mu}$.
The analysis carried out on the basis of hermitian realization
(\ref{5.2}) will also preserve all the results obtained so far,
particularly it will not change the relations (\ref{4.1}) and (\ref{4.2}),
except only for the modification
  introduced by the additional factors, $A(P)$ and $\tilde{A}(P,Q),$
\begin{equation} \label{5.4}
 e^{iP {\x}^{h}} \triangleright 1 ~ = ~ e^{i{\bf{K}}_{\mu}(P)x^{\mu}} A(P)
\end{equation}
and
\begin{equation} \label{5.5}
 e^{iP {\x}^{h}} \triangleright e^{iQx} ~ = ~ e^{i{\bf{P}}_{\mu}(P,Q)x^{\mu}} \tilde{A}(P,Q),
\end{equation}
where the unit element $1$ is defined in (\ref{2.19}) and the
quantities ${\bf{P}}_{\mu}(P,Q)$ and ${\bf{K}} (P)$ are given as
before, Eqs.(\ref{4.2a}) and (\ref{4.2b}), respectively.
These last two relations  can be inverted and rewritten as
\begin{eqnarray} \label{5.6}
 e^{i{\bf{K}}^{-1} (P) {\x}^{h}} \triangleright 1 ~ & = & ~ e^{i P_{\mu} x^{\mu}}
 A({\bf{K}}^{-1} (P)) \nonumber \\
 & \equiv & ~ e^{i P x} ~ \sqrt[4]{1+ a^2 P^2}
\end{eqnarray}
and
\begin{eqnarray} \label{5.7}
 e^{i{\bf{K}}^{-1} (P) {\x}^{h} } \triangleright e^{iQx} ~ & = & ~ e^{i{\bf{P}}_{\mu}({\bf{K}}^{-1} (P),Q)x^{\mu}}
  \tilde{A} ({\bf{K}}^{-1} (P),Q) \nonumber \\
 & \equiv & ~ e^{i{\mathcal{D}}(P,Q)x} ~ \frac{\sqrt[4]{1+ a^2 {({\mathcal{D}}(P,Q))}^2}}{\sqrt[4]{1+ a^2 Q^2}},
\end{eqnarray}
where ${\bf{K}}^{-1} (P)$ and ${\mathcal{D}}(P,Q)$ are given in (\ref{4.4}) and
(\ref{4.16}), respectively, with the same relationship between ${\bf{P}}({\bf{K}}^{-1} (P),Q)$ and 
${\mathcal{D}}(P,Q)$ as in (\ref{4.12}).
In the  above identities, last lines give the form of the additional
 factors,
\begin{equation} \label{5.8}
 A({\bf{K}}^{-1} (P))  ~ = ~ \sqrt[4]{1+ a^2 P^2},
\end{equation}
\begin{equation} \label{5.9}
  \tilde{A}({\bf{K}}^{-1} (P),Q) ~ = ~ \frac{\sqrt[4]{1+ a^2 {({\mathcal{D}}(P,Q))}^2}}{\sqrt[4]{1+ a^2 Q^2}}. 
\end{equation}
As is the case with the identities (\ref{4.1}) and (\ref{4.2}) which are rigorously
established for the realization (\ref{2.10}), 
the identities in Eqs.(\ref{5.6}) and (\ref{5.7}) are also rigorously
established, but for the
realization ${\x}^{h},$ Eq.(\ref{5.2}). They can relatively easily be
checked up to second order in the deformation parameter $a.$ To do
this, one has to insert (\ref{4.4}) and (\ref{5.2}) into l.h.s. of (\ref{5.6}) (or (\ref{5.7})),
make Taylor expansion and gather all terms up to second order in $a.$
             
    On the basis of these results, we can introduce noncommutative
    plane waves ${\hat{e}}^{+}_{P}$ with label $+$ as follows
\begin{equation} \label{5.10}
 {\hat{e}}^{+}_{P} ~ \equiv ~ \frac{e^{i{\bf{K}}^{-1} (P) {\x}^{h}
 }}{ \sqrt[4]{1+ a^2 P^2} }.
\end{equation} 
This allows us to dissociate a general noncommutative field
$\hat{\phi} ({\x}^{h})$ into elementary Fourier components that are represented by plane
waves (\ref{5.10}),
\begin{equation} \label{5.10a}
  \hat{\phi} ({\x}^{h}) ~ = ~ \int d^n P ~ \tilde{\phi} (P) ~
    {\hat{e}}^{+}_{P}  
   ~ = ~ \int d^n P ~ \tilde{\phi} (P) ~ \frac{e^{i{\bf{K}}^{-1} (P) {{\x}^{h}}
 }}{ \sqrt[4]{1+ a^2 P^2} }.
\end{equation} 
The above expansion is consistent with the $\Omega$-map (\ref{2.20}) and with the
expansion (\ref{4.19}) that can be applied to a general element of the
algebra ${\mathcal{A}}_{a},$  i.e. to an arbitrary commutative field $\phi (x),$
\begin{equation} \label{5.10b}
  \hat{\phi} ({\x}^{h}) \triangleright 1 ~ = ~ \int d^n P ~ \tilde{\phi} (P) ~ \frac{e^{i{\bf{K}}^{-1} (P) {\x}^{h}
 }}{ \sqrt[4]{1+ a^2 P^2} } \triangleright 1 ~ = ~ \int d^n P ~
 \tilde{\phi} (P) ~ e^{iPx}  ~ = ~ \phi (x).
\end{equation} 
One can further show that the antipode $S(P)$ has the following property
\begin{equation} \label{5.11}
 {\bf{K}}^{-1} (S(P)) = - {\bf{K}}^{-1} (P).
\end{equation} 
This property will show up as the missing link required for
 introducing an adjoint operation $\dagger$ in a proper and consistent way
and for making a correct and complete correspondence between
  algebras ${\mathcal{A}}_{a}$ and ${\hat{\mathcal{A}}}_{a}$.
In particular, it allows us to make one-to-one correspondence between
 hermitian conjugated noncommutative plane waves and hermitian
 conjugated commutative plane waves. Thus, it is crucial in building
a field theory on $\kappa$-Minkowski spacetime in an internally
 consistent way. Due to the fact that realization ${\x}^{h},$
 Eq.(\ref{5.2}), is hermitian, the plane wave (\ref{5.10}) will be an
 unitary operator, so that corresponding hermitian conjugation will lead to
\begin{equation} \label{5.12}
  \frac{{\left( e^{i{\bf{K}}^{-1} (P) {\x}^{h}
 } \right)}^{\dagger}}{ \sqrt[4]{1+ a^2 P^2} }
  ~ = ~ \frac{ e^{-i{\bf{K}}^{-1} (P) {\x}^{h}
  } }{ \sqrt[4]{1+ a^2 P^2} } 
  ~ = ~ \frac{ e^{i{\bf{K}}^{-1} (S(P)) {\x}^{h}
  } }{ \sqrt[4]{1+ a^2 P^2} },
\end{equation} 
where in the final step use has been made of the property (\ref{5.11}).
The stated result enables us to introduce noncommutative 
 plane waves ${\hat{e}}^{-}_{P}$ with label $-$,
\begin{equation} \label{5.13}
 {\hat{e}}^{-}_{P} ~ \equiv ~ {({\hat{e}}^{+}_{P})}^{\dagger} ~ = ~ \frac{e^{i{\bf{K}}^{-1} (S(P)) {\x}^{h}
 }}{ \sqrt[4]{1+ a^2 P^2} }.
\end{equation} 
 It is worthy to note that due to the property (\ref{3.10a}) of the
 antipode, the noncommutative plane wave ${\hat{e}}^{-}_{P}$ can be identified with 
the noncommutative plane wave ${\hat{e}}^{+}_{S(P)}$ with label $+$, whose momentum
 is obtained by the transformation $P \longmapsto S(P).$ Thus, we have
 ${\hat{e}}^{-}_{P} = {\hat{e}}^{+}_{S(P)}. $

According to (\ref{5.6}) and the property (\ref{3.10a}) of the antipode,
  the plane waves (\ref{5.10}) and (\ref{5.13}) act on the unit element
$1$ as 
\begin{equation} \label{5.14}
 {\hat{e}}^{+}_{P} ~ \triangleright 1 ~ = ~ e^{iPx}, \qquad 
   {\hat{e}}^{-}_{P} ~ \triangleright 1 ~ = ~ e^{iS(P)x},
\end{equation}
realizing in this way the following one-to-one correspondence
\begin{eqnarray} \label{5.15}
  \Omega: ~ {\mathcal{A}}_{a} \longrightarrow {\hat{\mathcal{A}}}_{a}   \qquad
     \mbox{with the property}  \qquad
     e^{iPx} & \longmapsto &  {\hat{e}}^{+}_{P}, \nonumber \\
    e^{iS(P)x} & \longmapsto  & {\hat{e}}^{-}_{P}.
\end{eqnarray}
With these correlations we are now as well in a position to draw a
definite and unique correspondence  between hermitian conjugated
elements of the algebras ${\mathcal{A}}_{a}$ and ${\hat{\mathcal{A}}}_{a}.$
In particular, we can take an adjoint of the expansion 
(\ref{5.10a}) by applying $\dagger$ to it and then, by utilising Eq.(\ref{5.6}), act upon the unit
element (\ref{2.19}) to obtain
\begin{equation} \label{5.10c}
  {\hat{\phi}}^{\dagger} ({\x}^{h}) \triangleright 1 ~ = ~
  \int d^n P ~ {\tilde{\phi}}^{\ast} (P) ~ \frac{e^{i{\bf{K}}^{-1} (S(P)) {\x}^{h}
 }}{ \sqrt[4]{1+ a^2 P^2} } \triangleright 1 ~ = ~ \int d^n P ~
 {\tilde{\phi}}^{\ast} (P) ~ e^{iS(P)x}  ~ = ~ {\phi}^{\dagger} (x),
\end{equation} 
in accordance with (\ref{4.22}).
This makes the $\Omega$-map, $\Omega : {\mathcal{A}}_{a} \longrightarrow
 {\hat{\mathcal{A}}}_{a}, $ fully characterized now in a sense that we now also
 know how to make a correspondence between hermitian conjugated elements
 of the algebras ${\mathcal{A}}_{a}$ and ${\hat{\mathcal{A}}}_{a},$ namely,
\begin{equation} \label{5.10d}
 \Omega : {\mathcal{A}}_{a} \longrightarrow
 {\hat{\mathcal{A}}}_{a},  \qquad  {\phi}^{\dagger}(x)  \longmapsto
 {{\hat{\phi}}}^{\dagger} ({\x}^{h}), \qquad \mbox{such that}  \qquad
 {\hat{\phi}}^{\dagger}({\x}^{h}) ~ \triangleright 1 ~ = ~ {\phi}^{\dagger} (x).
\end{equation}
It is important to note that this correspondance is possible only for the
hermitian realization,
$ ~ {\hat{x}}^{h}, ~ {({\hat{x}}^{h})}^{\dagger} = {\hat{x}}^{h}, $
 because the
notion of the adjoint operation $\dagger$ has a sense in that case.
In other words, it is only in that case when the hermitian
representation of noncommutative coordinates is used that
 the adjoint operation $\dagger$ can be introduced in a consistent way.

The one-to-one correspondence described above directly explaines and
justifies why the prescription (\ref{4.18}), made in the previous section, is correct 
and  in agreement with the $\Omega$-map, Eq.(\ref{2.20}), which provides the communication
between two different descriptions of the same physics, one in terms of
ordinary, commutative fields and coordinates and the other in terms of
noncommutative coordinates and fields. This communication
between two descriptions, while
realized through the isomorphism $\Omega,$ was still not fully
specified all until
the moment when it became clear how this correspondence should look
like in the case of hermitian conjugated fields. Up to that moment,
the prescription for mapping noncommutative fields  into
commutative ones
was generally known, but it was not known what would hermitian
conjugated noncommutative field transform into under the $\Omega$-map
and how would the hermitian conjugation (adjoint operation) look like
at all.
 To solve these
ambiguities, it was first necessary to introduce the notion of realization which is hermitian,
so that noncommutative plane wave can be treated as an unitary operator.
The second important point was to realize that there exists identity
of the form (\ref{5.11}),
whose significance shows up as a crucial one in the construction process.
These favourable circumstances enabled the complete specification of the isomorphic $\Omega$-map,
 fixed up all problems and inconsistencies that have existed before
 and encompassed the whole picture
in a neat way.

   It is now possible to introduce a star product (\ref{2.21}) corresponding to
   hermitian realization ${\x}^{h},$ Eq.(\ref{5.2}). We designate it
   with ${\star}_h.$ According to the general definition (\ref{2.21}) of the star product,
  we can write
\begin{eqnarray} \label{5.16}
  e^{iPx}~ {\star}_h ~ e^{iQx} ~ & = & ~ \frac{e^{i{\bf{K}}^{-1} (P) {\x}^{h}
 }}{ \sqrt[4]{1+ a^2 P^2} }
 \frac{e^{i{\bf{K}}^{-1} (Q) {\x}^{h}
 }}{ \sqrt[4]{1+ a^2 Q^2} }  ~ \triangleright 1 \nonumber \\
  & = & ~ e^{i{\mathcal{D}}(P,Q)x} ~ \frac{\sqrt[4]{1+ a^2
 {({\mathcal{D}}(P,Q))}^2}}{\sqrt[4]{1+ a^2 P^2} ~ \sqrt[4]{1+ a^2 Q^2}},
\end{eqnarray}
where we have successively applied  identities (\ref{5.4}) and (\ref{5.5}).
The corresponding star product between arbitrary two elements $f$ and
$g$ of the algebra ${\mathcal{A}}_{a}$ modifies accordingly,
\begin{equation} \label{newstarproduct}
(f \; {\star}_h \; g)(x)  =   \lim_{\substack{u \rightarrow x }}
 m \left ( e^{x^{\alpha} ( \triangle - {\triangle}_{0}) {\partial}_{\alpha} }
  \sqrt[4]{\frac{1- a^2
   \triangle ( {\p}^2)}{(1- a^2 ~ {\p}^2 \otimes 1) ~ (1-
   a^2 ~ 1 \otimes {\p}^2 )}} ~
  f(u) \otimes g(u) \right ), 
\end{equation}
where the homomorphic property of the coproduct $\triangle ({\p}_{\alpha}),$ Eq.(\ref{coproductmomentum}),
is utilised, namely, $\triangle ({\p}_{\alpha}) \triangle ({\p}^{\alpha}) = \triangle ({\p}^2).$
In this way, the nonhermitian version of the star product, Eq.(\ref{4.17}), is
replaced by the hermitian one, Eq.(\ref{newstarproduct}).


It can be noted that unlike the star product (\ref{4.17}), the star
 product (\ref{newstarproduct}), corresponding to hermitian realization 
  (${({\x}^{h})}^{\dagger} = {\x}^{h}$), 
  breaks translational invariance.
This can be seen by following the same steps that result with relation (\ref{4.38}). 
  One only has to take care of the cosequences of the hermitization
 process. These imply that instead of relations (\ref{4.35}),
one has to deal with the following ones
\begin{equation} \label{5.241}
   e^{i {\bf{K}}^{-1}(P) {\x}^h} \triangleright 1
 ~ = ~ A({\bf{K}}^{-1}(P)) e^{i Px}, \qquad  e^{i {\bf{K}}^{-1}(Q) {\x}^h} \triangleright 1
 ~ = ~ A({\bf{K}}^{-1}(Q)) e^{i Qx},
\end{equation}
where  $A({\bf{K}}^{-1}(P))$ and $\tilde{A}({\bf{K}}^{-1}(P),Q)$ in
the expression below
are determined by Eqs.(\ref{5.8}) and (\ref{5.9}), respectively.
Having that, translation $ \x_{\mu} \longrightarrow \x_{\mu}  + \hat{y}_{\mu} $ leads to
\begin{eqnarray} \label{5.242}
 e^{i P(x+y)} ~ {\star}_h ~ e^{i Q(x+y)} ~ & = & ~  \frac{e^{i {\bf{K}}^{-1}(P) ({\x}^h
    + {\hat{y}}^h)}}{{\bigg[ A({\bf{K}}^{-1}(P)) \bigg]}^2}  \frac{e^{i {\bf{K}}^{-1}(Q) ({\x}^h
    + {\hat{y}}^h)}}{{\bigg[ A({\bf{K}}^{-1}(Q)) \bigg]}^2}  \triangleright 1  \nonumber \\
    ~ & = & ~  \frac{1}{{\bigg[ A({\bf{K}}^{-1}(P)) \bigg]}^2} \frac{1}{ A({\bf{K}}^{-1}(Q))}
 e^{i {\bf{K}}^{-1}(P) {\hat{y}}^h} e^{i {\bf{K}}^{-1}(Q) {\hat{y}}^h} \triangleright e^{i{\mathcal{D}}(P,Q)x} \tilde{A}({\bf{K}}^{-1}(P),Q)  \nonumber \\
  ~ & = & ~  {\Bigg[ \frac{\tilde{A}({\bf{K}}^{-1}(P),Q)}{A({\bf{K}}^{-1}(P))} \Bigg]}^2
 e^{i{\mathcal{D}}(P,Q) (x + y)},
\end{eqnarray}
leading to a conclusion that the star product (\ref{newstarproduct}) is
not translation invariant. This is due to the appearance of an extra
factor on r.h.s. of Eq.(\ref{5.242}). However, due to some other properties
of the new star product (\ref{newstarproduct}), the free scalar field
action, that is constructed in terms of this new star product, will nevertheless be
translationally invariant. We shall further discuss the issues
related with translational invariance in
the concluding section. For the moment let us focus our attention on the
other important properties of the new star product.


With this aim, it is interesting to note
 that noncommutative plane waves with opposite labels
(one $+$, the other $-$ and vice versa) are orthonormal among themselves,
\begin{eqnarray} \label{5.17}
 \int d^n x ~ {\hat{e}}^{-}_{P} ~ {\hat{e}}^{+}_{Q} ~ \triangleright 1 ~ & = & ~
 \int d^n x ~ \frac{e^{{\left( i{\bf{K}}^{-1} (P) {\x}^{h} \right)}^{\dagger}
 }}{ \sqrt[4]{1+ a^2 P^2} }
 \frac{e^{i{\bf{K}}^{-1} (Q) {\x}^{h}
 }}{ \sqrt[4]{1+ a^2 Q^2} }  ~ \triangleright 1 \nonumber \\
  & = & \int d^n x ~ \frac{e^{ i{\bf{K}}^{-1} (S(P)) {\x}^{h}
 }}{ \sqrt[4]{1+ a^2 P^2} } e^{iQx}
 = \int d^n x ~ \frac{\tilde{A}({\bf{K}}^{-1} (S(P)),Q)}{\sqrt[4]{1+ a^2 P^2}} ~ e^{i{\mathcal{D}}(S(P),Q)x} \nonumber \\
  & = & ~ \frac{\tilde{A}({\bf{K}}^{-1} (S(P)),Q)}{\sqrt[4]{1+ a^2 P^2}}
 ~ {(2\pi)}^{n} {\delta}^{(n)} \left( {\mathcal{D}}(S(P),Q) \right) = ~ \frac{\tilde{A}({\bf{K}}^{-1} (S(P)),Q)}{\sqrt[4]{1+ a^2 P^2}}
  ~ {(2\pi)}^{n} ~ \sqrt{1+ a^2 P^2}  ~ {\delta}^{(n)} (P-Q) \nonumber \\
  & = & ~ \frac{\sqrt[4]{1+ a^2
 {({\mathcal{D}}(S(P),Q))}^2}}{\sqrt[4]{1+ a^2 P^2} ~ \sqrt[4]{1+ a^2 Q^2}}
      ~ {(2\pi)}^{n} ~ \sqrt{1+ a^2 P^2}  ~ {\delta}^{(n)} (P-Q)
  ~ = ~ {(2\pi)}^{n} ~ {\delta}^{(n)} (P-Q).
\end{eqnarray}
 In the second line here, Eqs.(\ref{5.7}) and (\ref{5.11}) have been
 used, while in the third line the
 result for the ${\delta}^{(n)}$-function (\ref{4.29}) was used.
The expression considered will be different from zero only for $~Q=P~$
 and since in this case we
 have $~{\mathcal{D}}(S(Q),Q) = 0,~$ due to the very definition of the antipode $S(P),$
all factors in the last line of (\ref{5.17}) will cancel each other,
 leading to the orthonormality property of noncommutative plane waves with opposite
 labels. On the basis of the definition (\ref{2.21}) and
  correspondence (\ref{5.15}), the integral in (\ref{5.17}) can be recognized and rewritten in
 terms of the star product, Eq.(\ref{5.16}), as follows
\begin{eqnarray} \label{5.18}
 \int d^n x ~ {\hat{e}}^{-}_{P} ~ {\hat{e}}^{+}_{Q} ~ \triangleright 1 ~ & = & ~
 \int d^n x ~ \frac{e^{{\left( i{\bf{K}}^{-1} (P) {\x}^{h} \right)}^{\dagger}
 }}{ \sqrt[4]{1+ a^2 P^2} }
 \frac{e^{i{\bf{K}}^{-1} (Q) {\x}^{h}
 }}{ \sqrt[4]{1+ a^2 Q^2} }  ~ \triangleright 1 \nonumber \\
  & = & \int d^n x ~ e^{iS(P)x} ~ {\star}_{h} ~ e^{iQx},
\end{eqnarray}
which, according to (\ref{5.17}), is equal to $\; {(2\pi)}^{n} ~ {\delta}^{(n)} (P-Q).~$
Note the difference of this result when compared to the similar one
obtained in the previous section for the star product in nonhermitian realization.
In that case the plane waves were not orthonormal to each other and the result was not so simple.
    By taking an adventage of the result (\ref{5.17}), we may now 
    reconsider the expression (\ref{4.21}),
\begin{align} \label{5.19}
 \int d^n x ~ {\psi }^{\dagger} ~ {\star}_{h} ~ \phi, 
\end{align}
but this time with the star product ${\star}_{h}$ corresponding to
    hermitian realization (\ref{5.2}) and rederive the corresponding
    mathematical identity as in the previous section and see if it is modified with respect to 
       the identity (\ref{4.31}) obtained there. Thus, in accordance with Eqs.(\ref{4.18}),(\ref{4.19})
 and (\ref{4.22}) and by making use of the just derived result (\ref{5.17}),
    i.e. the result
\begin{equation} \label{5.19a}
 \int d^n x ~ e^{iS(P)x} ~ {\star}_{h} ~ e^{iQx} ~ = ~ {(2\pi)}^{n} ~ {\delta}^{(n)} (P-Q),
\end{equation}
  we easily find
\begin{eqnarray} \label{5.20}
 \int d^n x ~ {\psi }^{\dagger} ~ {\star}_{h} ~ \phi & = &
   \int d^n P \int d^n Q ~ {\tilde{\psi}}^{\ast} (P)
 {\tilde{\phi}} (Q) ~ \int d^n x ~ e^{iS(P)x} ~ {\star}_{h} ~ e^{iQx} \nonumber \\
  & = & \int d^n P \int d^n Q ~ {\tilde{\psi}}^{\ast} (P)
 {\tilde{\phi}} (Q) {(2\pi)}^{n} ~ {\delta}^{(n)} (P-Q).
\end{eqnarray}
Utilising further the representation (\ref{4.30}) for
 $ {\delta}^{(n)} $-function, we finally get 
\begin{align} \label{5.21}
 \int d^n x ~ {\psi }^{\dagger} ~ {\star}_{h} ~ \phi
 ~ = ~ \int d^n x ~ {\psi }^{\ast} (x)  \phi (x),
\end{align}
where $*$ denotes the standard complex conjugation operation in the
undeformed algebra ${\mathcal{A}}$.
Thus we get a result that the star product for $\kappa$-Minkowski space, in a hermitian
realization (\ref{5.2}), can be replaced with the ordinary pointwise
multiplication, under the integration sign. This property is known to
hold for canonical type of noncommutativity, as for example for the
Groenewold-Moyal plane, but up to this moment, was not established and recognized in the
context of $\kappa$-Minkowski space.

 The fact
that under the integration sign we can simply drop out the star product and
replace it with an ordinary pointwise multiplication has for consequence
that the free massive scalar field action on $\kappa$-deformed
Minkowski space loses a nonlocal character and takes on an undeformed
shape, at least in the part with the mass term. However, before being able to give the
final form to the action (\ref{4.20}), with star product $\star$ replaced by
${\star}_{h},$ we have to know how star product of vector fields
behaves upon integration. In other words, we have to know the right
form for the partial integration formula valid on $\kappa$-Minkowski
spacetime and applied to the star product (\ref{newstarproduct}). By going
through the same steps  and by following 
the similar lines of reasoning which brought us to the result
(\ref{5.21}), we get the required partial integration formula
in the form
\begin{equation} \label{5.21b}
 \int d^n x ~ {({\p}_{\mu} \psi ) }^{\dagger} ~ {\star}_{h} ~ \phi
 ~ = ~ - \int d^n x ~ { \psi }^{\dagger} ~ {\star}_{h} ~
 {\p}_{\mu} \phi  ~ = ~ - \int d^n x ~ {\psi }^{\ast} (x) {\p}_{\mu} \phi (x).
\end{equation}
A variant of this formula, when derivative is not affected by the
adjoint operation, looks as
\begin{equation} \label{5.21c}
 \int d^n x ~ {\p}_{\mu} {\psi }^{\dagger} ~ {\star}_{h} ~ \phi
 ~ = ~ \int d^n x ~ { \psi }^{\dagger} ~ {\star}_{h} ~
 S({\p}_{\mu}) \phi  ~ = ~ \int d^n x ~ {\psi }^{\ast} (x) S({\p}_{\mu}) \phi (x),
\end{equation}
with $S(-i{\p}_{\mu}) = -iS({\p}_{\mu})$ being the antipode (\ref{3.8}) for translation generators.
Having properties (\ref{5.21}) and (\ref{5.21b}) of the star
product (\ref{newstarproduct}), we are now in a position to obtain a
final form for the free massive scalar field action, 
\begin{eqnarray} \label{5.22}
 S[\phi] & = & \frac{1}{2}\int d^n x ~ {(\p_{\mu}\phi)}^{\dagger} ~
 {\star}_{h} ~ \p^{\mu}\phi +
 \frac{m^2}{2}\int d^n x ~ {\phi }^{\dagger} ~ {\star}_{h} ~ \phi \nonumber \\
 & = & -\frac{1}{2}\int d^n x ~ {\phi}^{\ast}(x) ~
  \p_{\mu} \p^{\mu}\phi (x) +
 \frac{m^2}{2}\int d^n x ~ {\phi }^{\ast}(x) \phi (x) \nonumber \\
  & = & \frac{1}{2} \int d^n x ~
  \bigg[ {({\p}_{\mu}{ \phi })}^{\ast} 
     ({\p}^{\mu}\phi ) (x)
 + m^2 {\phi }^{\ast} (x)  \phi (x) \bigg],
\end{eqnarray}
showing that the equivalent theory on undeformed space is no more nonlocal and
actually keeps the same form as the initial noncommutative theory.
This result is an immediate consequence of properties of the star
product (\ref{newstarproduct}) corresponding to hermitian realization (\ref{5.2}).
Due to property (\ref{5.21}), the action (\ref{5.22}) appears to be
translationally invariant, despite being written in terms of the  star
product that breaks translation invariance.
It should also be noted that the form of Klein-Gordon operator in the final
expression for the action is usual, undeformed one, $ -\p_{\mu} \p^{\mu} + m^2$.
This is in contrast with the analysis carried out in
Refs.\cite{Dimitrijevic:2003wv} and \cite{Grosse:2005iz}
where a deformed Klein-Gordon operator $\frac{2}{a^2}(1- \sqrt{1- a^2
  {\p}^2}) + m^2$ was used.

In the case that the original action is defined as
\begin{equation} \label{5.22a}
 S[\phi] ~ = ~  \frac{1}{2}\int d^n x ~ \p_{\mu} {\phi}^{\dagger} ~
 {\star}_{h} ~ \p^{\mu}\phi +
 \frac{m^2}{2}\int d^n x ~ {\phi }^{\dagger} ~ {\star}_{h} ~ \phi,
\end{equation}
the transformation according to partial integration formula
(\ref{5.21c}) would lead to
\begin{eqnarray} \label{5.22b}
 S[\phi] & = & \frac{1}{2}\int d^n x ~ {\phi}^{\dagger} ~
 {\star}_{h} ~ S(\p_{\mu}) \p^{\mu}\phi +
 \frac{m^2}{2}\int d^n x ~ {\phi }^{\dagger} ~ {\star}_{h} ~ \phi \nonumber \\
  & = & \frac{1}{2} \int d^n x ~
  \bigg[ { \phi }^{\ast} (x) ~
    S({\p}_{\mu}) {\p}^{\mu}\phi  (x)
 + m^2 {\phi }^{\ast} (x)  \phi (x) \bigg],
\end{eqnarray}
and the final form for the action would represent
an equivalent free massive scalar field theory on ordinary Minkowski spacetime,
with modified Klein-Gordon operator, $~S(\p_{\mu}) \p^{\mu} + m^2.$

In order to discuss the issue of reality of the scalar field $\phi (x),$
let us see what is happening if we change the variables of integration
in momentum space according to $P \; \longmapsto S(P). \;$ The
meassure in momentum space will then transform according to
\begin{equation} \label{5.23}
 d^n S(P) ~ = ~ \det\left( \frac{\p S(P_{\mu})}{\p P_{\nu}} \right)  d^n P.
\end{equation}
The Jacobian in (\ref{5.23}) needs to be found and
 in order to do this, in the same way as we have already done it before to calculate
 Jacobian (\ref{Jacobian}), we orient deformation
four-vector $a$ to point along the time-direction, $a=(a_0,0,...,0),$ in which case
the required matrix entries of the corresponding Jacobian of
transformation look as
\begin{eqnarray} \label{5.24}
  \frac{\p S(P_0)}{\p P_0} ~ & = & ~ -1 -a_0^2 {\vec{P}}^2
 \frac{Z(P)}{\sqrt{1- a_0^2 P^2}},  \nonumber \\
 \frac{\p S(P_0)}{\p P_i} ~ & = & ~ -a_0 P_i Z(P) \left( 2 + 
 \frac{a_0^2 {\vec{P}}^2}{\sqrt{1- a_0^2 P^2}} Z(P) \right),  \nonumber \\
 \frac{\p S(P_i)}{\p P_0} ~ & = & ~ - P_i
 \frac{a_0}{\sqrt{1- a_0^2 P^2}} Z(P), \nonumber \\
  \frac{\p S(P_i)}{\p P_j} ~ & = & ~ - Z(P) \left( {\delta}_{ij} + a_0^2
 \frac{P_i P_j}{\sqrt{1- a_0^2 P^2}} Z(P) \right).
\end{eqnarray}
Thus, the Jacobian alone is given by
\begin{equation} \label{Jacobian2}
 \det \left( \frac{\p S(P_{\mu})}{\p P_{\nu}} \right) =
\left| \begin{array}{ccccc}
  -1-a_0 {\vec{P}}^2 \tilde{b} &  -a_{0} P_1 Z(P) \left( 2+ a_0 {\vec{P}}^2 \tilde{b} \right)  & 
   \cdot \cdot \cdot & -a_{0} P_{n-2} Z(P) \left( 2+ a_0 {\vec{P}}^2 \tilde{b} \right)
  & ~  -a_{0} P_{n-1} Z(P) \left( 2+ a_0 {\vec{P}}^2 \tilde{b} \right) \\
 -\tilde{b} P_1   & - Z(P) \left( 1+a_{0} \tilde{b} P_{1}^2 \right) &  \cdot \cdot \cdot
 & - a_{0} \tilde{b} Z(P) P_{1} P_{n-2} & - a_{0} \tilde{b} Z(P) P_{1} P_{n-1} \\ 
 \cdot & \cdot &   \cdot \cdot \cdot & \cdot & \cdot \\
\cdot & \cdot &  \cdot \cdot \cdot & \cdot & \cdot \\
 -\tilde{b} P_{n-3}  &  - a_{0} \tilde{b} Z(P) P_{n-3} P_{1} &  \cdot \cdot \cdot 
 & - a_{0} \tilde{b} Z(P) P_{n-3} P_{n-2} & - a_{0} \tilde{b} Z(P) P_{n-3} P_{n-1} \\
  -\tilde{b} P_{n-2} & - a_{0} \tilde{b} Z(P) P_{n-2} P_{1} & \cdot \cdot \cdot
 & - Z(P) \left( 1+ a_{0} \tilde{b} P_{n-2}^2 \right) & - a_{0} \tilde{b} Z(P) P_{n-2} P_{n-1} \\
  -\tilde{b} P_{n-1}  & - a_{0} \tilde{b} Z(P) P_{n-1} P_{1} & \cdot \cdot \cdot 
 & - a_{0} \tilde{b} Z(P) P_{n-1} P_{n-2} & - Z(P) \left( 1+ a_{0} \tilde{b} P_{n-1}^2 \right) \\
\end{array} \right|
\end{equation}
where the  quantity $\tilde{b}$ is defined as
\begin{equation} \label{5.25}
  \tilde{b} ~ = ~ \frac{a_0}{\sqrt{1- a_{0}^2 P^2}} Z(P).
\end{equation}
The Jacobian (\ref{Jacobian2}) can be calculated to give
\begin{equation} \label{5.26}
 d^n S(P) ~ = ~ Z^{n-1}(P) d^n P,
\end{equation}
with $Z(P)$ given in (\ref{4.13}).

   The commutative scalar field $\phi (x)$ will  be real if it
   satisfies the condition
\begin{equation} \label{5.27}
 {\phi}^{\dagger} (x)  ~ = ~ {\phi} (x).
\end{equation}
From this condition, from the expansions (\ref{4.19})
 and (\ref{4.22}) and from Eq.(\ref{5.26}), we find that the
 reality condition (\ref{5.27}) imposed on the scalar field,
\begin{eqnarray} \label{5.28}
 {\phi}^{\dagger} (x)  ~ & =&  ~ \int d^n P ~ {\tilde{\phi}}^{\ast} (P) ~
 e^{iS(P)x} \nonumber \\
 & = & ~ \int d^n S(P) ~Z^{-(n-1)}(P) ~ {\tilde{\phi}}^{\ast} (P) ~ e^{iS(P)x}  ~ =  ~
 \int d^n S(P) ~ {\tilde{\phi}} (S(P)) ~ e^{iS(P)x}  ~ = ~ {\phi} (x),
\end{eqnarray}
leads to the following condition involving its Fourier components,
 \begin{equation} \label{5.29}
 {\tilde{\phi}} (S(P)) ~ = ~ Z^{-(n-1)} (P) {\tilde{\phi}}^{\ast} (P).
\end{equation}
The same condition would emerge if we imposed the reality condition on
the noncommutative fields, i.e. ${\hat{\phi}}^{\dagger}({\x}^{h}) = {\hat{\phi}}({\x}^{h}). $

\section{Concluding remarks and discussion}

We now turn again to discussion of the properties of the star product (\ref{newstarproduct})
corresponding to hermitian realization (\ref{5.2}).
With this purpose, the identity (\ref{5.21}) can successively be
applied twice to get
\begin{align} \label{5.21a}
 \int d^n x ~ {\psi }^{\dagger} ~ {\star}_{h} ~ \phi
 ~ = ~ \int d^n x ~  {({\phi}^{\ast} (x))}^{\ast} {\psi }^{\ast} (x) 
 ~ = ~ \int d^n x ~ {{\phi}^{\ast}}^{\dagger}  ~ {\star}_{h} ~  {\psi }^{\ast},
\end{align}
showing that the star product (\ref{newstarproduct}) has a generalized
trace property.
We point out that in the approach presented here, the generalized trace and cyclic properties
arise in a purely natural way, without having to make any artificial
interventions by hand. In Refs.\cite{Dimitrijevic:2003wv},\cite{Dimitrijevic:2003pn},\cite{Moller:2004sk}
an attempt was made to get trace and cyclic properties, satisfied by an integral defined on
$\kappa$-Minkowski spacetime. This attempt consisted in finding an
appropriate integration meassure which would enable integral to have
desired properties. However, described procedure appeared to have few
stumbling blocks and did not solve the problem in a completely
satisfactory way. In the approach carried out in this paper, the generalized trace
and cyclic properties instead emerge quite naturally, simply by demanding that the classical Dirac
operator representation (\ref{2.10}) has to be hermitian, giving rise
to hermitian realization (\ref{5.2}). The star product corresponding
to this realization then happens to have these  nice properties,
required for building any gauge theory.

   It is obvious from relation (\ref{5.21a}) that the integral will
   exhibit standard trace property, provided that classical physical
   fields $\psi$ and $\phi$ satisfy the conditions ${{\psi}^{\ast}}^{\dagger} = \psi$ 
  and ${{\phi}^{\ast}}^{\dagger} = \phi$. In this case we would thus have
\begin{align} \label{traceproperty}
 \int d^n x ~ {\psi }^{\dagger} ~ {\star}_{h} ~ \phi
 ~ = ~ \int d^n x ~ \phi ~ {\star}_{h} ~  {\psi }^{\dagger}.
\end{align}
In relation (\ref{5.19a}) we have calculated the corresponding integral
for one definite order of plane wave factors. If the plane wave factors within
this integral are reversed, the result would be
\begin{equation} \label{trace1}
 \int d^n x ~ e^{iQx} ~ {\star}_{h} ~ e^{iS(P)x}  ~ = ~ {(2\pi)}^{n} ~ {\delta}^{(n)} (S(P)-S(Q)).
\end{equation}
The right hand side of Eq.(\ref{trace1}) can be calculated with the
help of the known result for the Jacobian (\ref{Jacobian2}) to give
\begin{equation} \label{trace2}
 {\delta}^{(n)} (S(P)-S(Q))  ~ = ~ Z^{-(n-1)}(P) ~ {\delta}^{(n)} (P-Q).
\end{equation}
This would consequently lead to a generalized trace property at the
level of plane waves,
\begin{equation} \label{trace3}
 \int d^n x ~ e^{iS(P)x}~ {\star}_{h} ~e^{iQx} ~ = ~ Z^{n-1}(P) ~ \int d^n x ~ e^{iQx} ~ {\star}_{h} ~ e^{iS(P)x},
\end{equation}
where $Z^{n-1}(P)$ is a function of $P,$ the same one that appears in (\ref{5.26}).
If the use is further made of Fourier expansions (\ref{4.19}) and (\ref{4.22}),
and $Z^{n-1}(P)$ is transcribed into a form of differential operator
(by simply setting $P_{\mu} = -i{\p}_{\mu}$),
then Eq.(\ref{trace3}) would imply an alternative form of the identity
embracing the generalized trace property,
\begin{align} \label{trace4}
 \int d^n x ~ {\psi }^{\dagger} ~ {\star}_{h} ~ \phi
 ~ = ~ \int d^n x ~ (Z^{n-1} \phi) ~ {\star}_{h} ~  {\psi }^{\dagger}
  ~ = ~ \int d^n x ~  \phi ~ {\star}_{h} ~  Z^{-(n-1)} {\psi }^{\dagger}  ,
\end{align}
where $Z^{n-1}$ is now a differential operator, with $Z^{-1}$ given in (\ref{16}).
 The expression to the most right in relation (\ref{trace4}) is a
 direct  consequence of the property
 (\ref{3.10c}) of the antipode for translation generators.

As promised before, we turn to discussion regarding the issue of
translation invariance for theories defined on $\kappa$-spaces.
 With this purpose, we recall that the star product (\ref{4.11}), i.e. (\ref{4.17}),
 with the hermitization procedure not being implemented in, is
 translationally invariant. This we inferred by following the
 arguments made in Refs.\cite{Kosinski:1999dw},\cite{Daszkiewicz:2007eq}
and by invoking that
 $\kappa$-Minkowski space is invariant \cite{Daszkiewicz:2007eq} under translations, 
$~ \x_{\mu} \longrightarrow \x_{\mu}  + \hat{y}_{\mu}, $
provided that following conditions are satisfied:
\begin{equation} 
 [ \hat{y}_{\mu}, \hat{y}_{\nu}] = i(a_\mu \hat{y}_{\nu} - a_\nu \hat{y}_{\mu}),
 \qquad [ \x_{\mu}, \hat{y}_{\nu}] = 0.
\end{equation}
Using given arguments we were able to show that
 the star product in nonhermitian realization (see Eq.(\ref{2.10})) is translationally invariant,
\begin{equation} 
 e^{i P(x+y)} ~ \star ~ e^{i Q(x+y)} ~  =  ~ e^{i{\mathcal{D}}(P,Q) (x + y)}.
\end{equation}

On the other side, in the case of the star product (\ref{5.16}), i.e. (\ref{newstarproduct}) 
in hermitian realization (see Eq.(\ref{5.2})), translation $~ \x_{\mu}
\longrightarrow \x_{\mu}  + \hat{y}_{\mu} $ has led us to the conclusion
\begin{equation} \label{reply1}
 e^{i P(x+y)} ~ {\star}_h ~ e^{i Q(x+y)} ~  =  ~
    {\Bigg[ \frac{\tilde{A}({\bf{K}}^{-1}(P),Q)}{A({\bf{K}}^{-1}(P))} \Bigg]}^2
 e^{i{\mathcal{D}}(P,Q) (x + y)},
\end{equation}
showing that a redundant extra factor appears on the r.h.s. of Eq.(\ref{reply1}), leading
to a breaking of translational symmetry. This shows that
 the hermitization process on
 $\kappa$-Minkowski in some sence interferes with translation
 symmetry, at least in the setting given by the classical Dirac operator representation
(\ref{2.12}).
Few comments are in order, regarding translation symmetry breaking:
\begin{enumerate}
\item 
    It is evident  from Eq.(\ref{reply1}) that the condition for having
   translational invariance should be
 \begin{equation} \label{reply2}
 A({\bf{K}}^{-1}(P)) = 1.
 \end{equation}
   For the classical Dirac operator representation, i.e. that one characterized by
(\ref{2.12}), the condition (\ref{reply2}) is obviously not satisfied.
Thus, we have to conclude that classical Dirac operator
  realization, Eqs.(\ref{2.10}),(\ref{2.12}), is the one in which it is not possible to simultaneously 
  have translation invariance property along with hermiticity.
  However, it does not mean that it is not possible to find
  realizations, even a whole family of realizations for which these
  two important requirements do not interfere. For example, in
  Ref.\cite{Borowiec:2008uj},
   the authors consider certain family of realizations which are
  hermitian and because satisfying the condition $[x_\mu, \Phi_{\nu
  \lambda} (\p)] =0$,
  have this $A({\bf{K}}^{-1}(P))$ factor equal to $1$ and consequently the star
  product constructed in that realizations will have the required translation
  invariance property, along with hermiticity. 
   The analysis of these realizations, their 
  Hopf algebraic descriptions and corresponding star products would be
  an interesting subject for future investigations.
\item  
   In this paper we are considering a free scalar field theory whose action
   includes products of two fields.
      Because of the fundamental property (\ref{5.21}) of the star product
   ${\star}_h$, established in this paper (removing a star product under integration sign),  
    in the case of free scalar field theory we don't even in principle have a problem with
   translation invariance breaking.
\item 

  In case of the interacting field theory, it seems that the breaking of
  translational symmetry is unavoidable. However, it is not clear at
  all what are the physical consequences of translational symmetry breaking.
   These consequences should be carefully investigated, especially the possible violation of
   energy-momentum conservation \cite{Balachandran:2007sh}. Even if
  this situation with translational symmetry breaking could cause violation of
   energy-momentum conservation, we point out that in the
  case of the star product (\ref{newstarproduct}), this violation would be so minute,
   that one could safely neglect it. This conclusion can be drawn from
  the form of the factor
   $A({\bf{K}}^{-1}(P)), $ whose lowest order corrections to its value are of
  the second order in deformation parameter $a,$
 $$
 A({\bf{K}}^{-1}(P)) = 1 + {\mathcal{O}} (a^2).
 $$ 
Thus, if $a =\frac{1}{\kappa}$ is taken to correspond to Planck length
(where $\kappa$ is the Planck mass), then $a^2$ contributions are really of a very small magnitude.
  If one would set up to analyse an interacting field theory on $\kappa$-space, constructed
  from the star product (\ref{newstarproduct}) in hermitian realization, then he would not
  encounter even this minute signal of energy-momentum conservation
  violation in case  that he
  restricts his calculations  to a first order corrections in deformation parameter $a$.
Thus, when considering the interacting field theory on $\kappa$-space
constructed in terms of star product (\ref{newstarproduct}), we do not expect
a violation of energy-momentum conservation to show up within the first order in deformation parameter $a$.

\end{enumerate} 


As far as the issue of Lorentz invariance of the field theory on 
$\kappa$-Minkowski space is concerned, there is no definite answer yet
in a response to this question.
In most literature on the subject of $\kappa$-deformation, the
deformation vector $a_{\mu}$ is treated as a vector that obeys standard rules
for raising and lowering indices, realized through the application of the metric tensor,
but anyway its components are treated as constant parameters in every Lorentz frame.
We point out that such view inevitably leads to Lorentz violation.
In reference \cite{Freidel:2007hk} the authors also noticed this
problem. Their actual momentum space for field theory was determined by the region in
de Sitter space which is defined by the condition that is not Lorentz invariant.
Following these conclusions they argue that it is to expect that any theory
  with such momentum space will suffer from Lorentz violation.
Anyway, there has been an attempt made \cite{Arzano:2009ci} in order to overcome 
this problem which, according to Ref.\cite{Arzano:2009ci},
was successfully solved by modifying the action 
 of Lorentz generators. This modification, as claimed in \cite{Arzano:2009ci}, avoids the problem of Lorentz breaking.

However, we want to stress that in our approach the only possible way
in which Lorentz symmetry can be restored is to treat deformation
vector $a_{\mu}$ as a pure $n$-vector, which, besides obeying raising
and lowering of indices by means of the metric tensor, indeed transforms as a
$n$-vector under Lorentz transformations,   
\begin{equation} \label{5.30}
[M_{\mu\nu}, a_{\lambda}] ~ = ~ a_{\mu} \eta_{\nu\lambda} - a_{\nu}
  \eta_{\mu\lambda}.
\end{equation}
Here we assume that ${\x}_{\mu}$ and $a_{\nu}$ commute among
themselves, $[{\x}_{\mu}, a_{\nu}] = [a_{\mu}, a_{\nu}] = 0.$
This would imply that NC coordinates also transform as a $n$-vector under Lorentz transformations,
so that instead of relation (\ref{2.3}), we would now have
\begin{equation} \label{5.31}
[M_{\mu\nu}, \x_{\lambda}] ~ = ~ \x_{\mu} \eta_{\nu\lambda} - \x_{\nu}
  \eta_{\mu\lambda},
\end{equation}
with all other algebraic relations between the generators $\x_{\mu}, ~
M_{\mu \nu}$ and $P_{\mu}$ remaining unchanged \cite{KresicJuric:2007nh}.
If we are about to keep the representation (\ref{2.10}) (or
(\ref{5.2}) in the case of hermitian realization) for NC coordinates intact,
 then the algebraic setting just described requires a modification of
representation (\ref{2.8}) for the Lorentz generators, so that a new
form they acquire would look as
\begin{equation}
 M_{\mu \nu} = x_{\mu}\p_{\nu}- x_{\nu}\p_{\mu} + a_{\mu} \frac{\p}{\p a_{\nu}} - 
   a_{\nu} \frac{\p}{\p a_{\mu}}.
\end{equation}
 The relation (\ref{5.31}), comprising vector-like properties
 of NC coordinates, would have a far reaching consequence on the
 coalgebraic sector of the Hopf $\kappa$-Poincar\'{e} algebra.  Particularly, it would affect a
Lorentz part of the coalgebra by greatly simplifying it, giving rise
to an undeformed coproduct for Lorentz generators,
\begin{equation}
 \triangle M_{\mu \nu} ~ = ~ M_{\mu \nu} \otimes 1 + 1 \otimes  M_{\mu \nu},
\end{equation}
replacing the relation (\ref{coproductangmomentum}). On the other
hand, the coproduct for translation generators would remain unchanged,
still given by the relation (\ref{coproductmomentum}).
In this case the symmetry properties of the field theory
constructed in a described setting, would be enclosed by a Hopf
algebra whose Lorentz symmetry is undeformed at both algebraic and
coalgebraic level. Similar property is observed in Refs.\cite{Battisti:2008xy},\cite{Battisti:2010sr} in the context of
Snyder spacetime.
 All deformations then would be encoded within the coproduct 
(\ref{coproductmomentum}) for translation generators only, which in
particular is coassociative. The corresponding star product (\ref{newstarproduct})
would consequently be associative, with homomorphism relating
corresponding structures. Similar ideas concerning Lorentz covariance
are presented in Ref.\cite{Dabrowski:2009mw}, where the authors
consider Lorentz covariant $\kappa$-Minkowski spacetime.
 However, to find out whether Lorentz
symmetry is really preserved
or is actually broken below the Planck scale, we
 should wait for an experiment to get right and definite answer.


In this paper we have presented a construction of the star product on
noncommutative  $\kappa$-Minkowski spacetime in the setting provided
by the classical Dirac operator representation. The
 construction alone relies on the very property of hermiticity and
 based on this property
 directly leads to a formulation of invariant integral having generalized trace property. 
 This property is essential for building any gauge theory. To our
 knowledge, it is for the first time in the literature that 
 invariant integral on $\kappa$-deformed space, having properties
 (\ref{5.21}) and (\ref{5.21a}), has been constructed. 
However, obtained star product is not translationally invariant,
showing internal incompatibility between properties of translation
symmetry and hermiticity on $\kappa$-deformed spaces, at least for the
classical Dirac operator representation.


\noindent
{\bf Acknowledgment}\\
 This work was supported by the Ministry of Science and Technology of
 the Republic of Croatia under contract No. 098-0000000-2865. 


\end{document}